\documentclass[letterpaper,titlepage,11pt]{article}

\usepackage[colorlinks=true,urlcolor=blue,citecolor=magenta]{hyperref}
\usepackage{amssymb,amsmath,amsfonts}
\usepackage{cancel}
\usepackage{comment}
\usepackage{epsfig}
\usepackage{graphicx}
\usepackage{epstopdf}
\usepackage{caption}
\usepackage{subcaption}
\usepackage{amscd}
\usepackage{amsthm}
\usepackage{latexsym}
\usepackage{cite}
\usepackage{amsbsy}
\usepackage{bbm}
\usepackage[english]{babel}
\usepackage{psfrag}
\usepackage{tabularx}
\usepackage[normalem]{ulem}

\usepackage{overpic}
\usepackage{array}
\usepackage{rotating}
\usepackage{lipsum}
\usepackage{color}

\def\be{\begin{equation}}
\def\ee{\end{equation}}
\def\bea{\begin{eqnarray}}
\def\eea{\end{eqnarray}}

\allowdisplaybreaks

\newcommand{\Lieg}{\mathfrak{g}}
\newcommand{\GroupG}{\mathcal{G}}
\newcommand{\gel}{T}
\newcommand{\Gel}{g}
\newcommand{\gione}{A}

\newcommand{\Lieh}{\mathfrak{h}}
\newcommand{\GroupH}{\mathcal{H}}
\newcommand{\hel}{T}
\newcommand{\Hel}{h}
\newcommand{\hione}{I}

\newcommand{\cosetm}{\mathfrak{m}}
\newcommand{\cosetM}{\mathcal{M}}
\newcommand{\mel}{T}
\newcommand{\mione}{a}
\newcommand{\mitwo}{b}
\newcommand{\mithree}{c}
\newcommand{\mifour}{d}
\newcommand{\miPone}{i}
\newcommand{\miPtwo}{j}
\newcommand{\miPthree}{k}
\newcommand{\miPfour}{\ell}

\newcommand{\Span}{\textrm{span}}

\newcommand{\MCF}{\mathcal{A}}

\newcommand{\viel}{\mathbf{e}}

\newcommand{\gaugef}{m}

\newcommand{\lgb}{\lambda}

\newcommand{\lut}{\sigma}

\newcommand{\lgbm}{\zeta}
\newcommand{\lgbe}{\xi}

\newcommand{\sech}{\textrm{sech}}

\newcommand{\sct}{\omega}

\newcommand{\Lie}{\mathcal{L}}

\newcommand{\Kill}{\mathcal{K}}

\newcommand{\qA}{A}
\newcommand{\qB}{B}

\newcommand{\qc}{c}

\newcommand{\qJ}{J}
\newcommand{\qG}{G}
\newcommand{\qC}{C}
\newcommand{\qP}{P}
\newcommand{\qH}{H}
\newcommand{\qN}{N}
\newcommand{\qD}{D}

\newcommand{\qK}{K}

\newcommand{\qrJ}{M}
\newcommand{\qrP}{\Pi}

\newcommand{\qg}{\mathfrak{g}}
\newcommand{\qh}{\mathfrak{h}}
\newcommand{\qm}{\mathfrak{m}}

\newcommand{\qu}{u}

\usepackage[normalem]{ulem}

\def\clock{{\count0=\time
           \divide\count0 60
           \ifnum\count0<10 0\fi\the\count0
           \multiply\count0 -60 \advance\count0 \time
           :\ifnum\count0<10 0\fi \the\count0
         }}
\newcommand{\timestamp}{{\small\vbox{\hbox{\tt\jobname.tex}
\hbox{\the\day/\the\month/\the\year, \clock}}}}

\numberwithin{equation}{section}

\newcommand{\Nrel}{Nonrelativistic }
\newcommand{\nrel}{nonrelativistic }

\newcommand{\ndeg}{non-degenerate }
\newcommand{\ndegafternoun}{nondegenerate }

\begin{document}

\begin{titlepage}
\rightline{\vbox{   \phantom{ghost} }}

 \vskip 1.8 cm
\begin{center}
{\huge \bf
Homogeneous \Nrel Geometries as Coset Spaces}
\end{center}
\vskip .5cm

\centerline{\large {{\bf Kevin T. Grosvenor$^1$, Jelle Hartong$^2$, Cynthia Keeler$^3$, Niels A. Obers$^1$}}}

\vskip 1.0cm

\begin{center}

\sl $^1$ The Niels Bohr Institute, Copenhagen University,\\
\sl  Blegdamsvej 17, DK-2100 Copenhagen \O , Denmark\\
\sl $^2$ Institute for Theoretical Physics and Delta Institute for Theoretical Physics,\\ 
University of Amsterdam, Science Park 904, 1098 XH Amsterdam, The Netherlands\\
\sl $^3$  {Physics Department, Arizona State University, Tempe, AZ 85287, USA} \\ 
\vskip 0.4cm

\end{center}

\vskip 1.3cm \centerline{\bf Abstract} \vskip 0.2cm \noindent

We generalize the coset procedure of homogeneous spacetimes in (pseudo-)Riemannian geometry to non-Lorentzian geometries. These are manifolds endowed with nowhere vanishing invertible vielbeins that transform under local non-Lorentzian tangent space transformations. In particular we focus on \nrel symmetry algebras that give rise to (torsional) Newton-Cartan geometries, for which we demonstrate how the Newton-Cartan metric complex is determined by degenerate co- and contravariant symmetric bilinear forms on the coset. In specific cases we also show the connection of the resulting \nrel coset spacetimes to pseudo-Riemannian cosets via In\"on\"u-Wigner contraction of relativistic algebras as well as null reduction. Our construction is of use for example when considering limits of the AdS/CFT correspondence in which \nrel spacetimes appear as gravitational backgrounds for \nrel string or gravity theories.

\vspace{2pc}

\end{titlepage}

\pagestyle{empty}
\tableofcontents
\normalsize
\newpage
\pagestyle{plain}
\setcounter{page}{1}






\section{Introduction}

The interplay between geometry and symmetries plays a central role in physics and mathematics. 
As a prominent example, one can obtain (pseudo)-Riemannian geometry
by gauging the Poincar\'e algebra and turning spacetime translations into diffeomorphisms leaving the Lorentz transformations as the
internal, tangent space, gauge symmetries. Correspondingly, a particular useful class of pseudo-Riemannian spacetimes are homogeneous spacetimes
which can be obtained via a coset construction.  For example, Minkowski spacetime can be seen as the (left) coset space 
$\GroupG / \GroupH$, in which
$\GroupG$ is the Poincar\'e group and $\GroupH$ the subgroup corresponding to Lorentz transformations. 
Recently, non-Lorentzian geometry, with Newton-Cartan geometry as an important case, has gained interest due to its applications
in non-AdS holography \cite{Christensen:2013lma,Christensen:2013rfa,Hartong:2014oma,Hofman:2014loa,Hartong:2015usd,Jensen:2017tnb}, field theory
 \cite{Son:2013rqa,Geracie:2014nka,Jensen:2014aia,Hartong:2014pma,Hartong:2015wxa,Geracie:2015xfa,Gromov:2015fda,Fuini:2015yva,Bergshoeff:2015sic,Geracie:2016inm,Gromov:2017qeb,Arav:2016xjc,Auzzi:2017wwc}, 
 gravity \cite{Bergshoeff:2015uaa,Hartong:2015zia,Hartong:2015xda,Afshar:2015aku,Bergshoeff:2016lwr,Hartong:2016yrf,Bergshoeff:2017btm,VandenBleeken:2017rij,Bergshoeff:2017dqq} and string theory \cite{Andringa:2012uz,Harmark:2017rpg}. The present paper has two aims: first, to address the  natural question whether the notion of coset spacetime can be extended to these more general geometries; second, to present a formulation that can be of use when considering \nrel spaces as possible gravitational backgrounds dual to \nrel field theories.

Just as in the (pseudo)-Riemannian case, a non-Lorentzian geometry begins with some  spacetime symmetry group, containing at least time and space translations as well as spatial rotations (isotropy),
but no Lorentz boosts.  Again, after gauging the translations, one obtains the local tangent space gauge symmetries from the local rotations and other possible symmetries, e.g. local \nrel or ultra-relativistic boost symmetry (see e.g. \cite{Andringa:2010it,Bergshoeff:2014uea,Hartong:2015zia,Hartong:2015xda,Festuccia:2016awg}). 
Examples are Newton--Cartan and Carrollian geometries whose tangent space structure is dictated by the Bargmann (centrally extended Galilei) and Carroll (zero speed of light contraction of Poincar\'e) algebras. Thus, a  non-Lorentzian geometry is a manifold that 
is locally flat in the sense of a kinematical principle of relativity that is different from Einstein's equivalence principle. Put another way, they are manifolds endowed with nowhere vanishing invertible vielbeins that transform under local non-Lorentzian tangent space transformations. In this terminology, non-Lorentzian theories of gravity are those that are obtained by making non-Lorentzian geometries dynamical. 

Beside its intrinsic mathematical relevance, our  study is motivated by the fact that having a coset description of a spacetime is
a powerful tool to implement symmetries of the geometry and find solutions to theories of gravity. A coset picture also aids in understanding geodesics and thus particle dynamics. On the mathematical side, our results are relevant for the classification of the \nrel analogue of maximally symmetric spacetimes and for finding solutions of e.g. Ho\v{r}ava-Lifshitz gravity or Carrollian gravity. Moreover, in view of employing
Ho\v{r}ava-Lifshitz gravity \cite{Griffin:2012qx,Janiszewski:2012nf}  or other non-Lorentzian gravity theories \cite{Hartong:2016yrf}
as bulk theories in holography,  non-Lorentzian coset backgrounds could play an important role as alternative spacetime duals of \nrel field theories, 
in the same way that AdS can be thought of as the largest homogeneous space from a coset of the conformal group exhibited in the dual CFT.
Furthermore, in the context of  probes of non-Lorentzian spacetimes, having a coset description 
could provide important insights towards a generalization of the Ryu--Takayanagi \cite{Ryu:2006bv} prescription of holographic entanglement%
\footnote{See e.g. Refs. \cite{Ammon:2013hba,deBoer:2013vca,Castro:2015csg,Song:2016pwx,Song:2016gtd} and \cite{Jiang:2017ecm} for the generalization of holographic entanglement entropy to  the case of higher spin theories and warped CFTs respectively.} 
 in non-Lorentzian holographic bulk duals. 

In view of their importance in \nrel holography%
\footnote{See \cite{Taylor:2015glc} for a review on Lifshitz holography}, Lifshitz and Schr\"odinger spacetimes were already studied via coset constructions in  \cite{SchaferNameki:2009xr} (and applied in e.g. \cite{Jottar:2010vp,Bagchi:2010xw,Duval:2012qr}). 
Since \nrel algebras
are not semi-simple and have a degenerate Cartan-Killing metric, the standard (Riemannian) coset method of 
contracting the (inverse) vielbeins with the (inverse) Cartan-Killing metric, obtaining the metric and its inverse, does not apply.
Instead of the Cartan-Killing metric,  \cite{SchaferNameki:2009xr} uses the most general  \ndeg group invariant symmetric bilinear form, as \cite{Nappi:1993ie} did for the Nappi-Witten  WZW model on the non-semi simple group $E_2^c$, which is the centrally extended 2-dimensional Euclidean group. Thus \cite{SchaferNameki:2009xr} recovers the so-called Lifshitz and Schr\"odinger spacetimes, first proposed in \cite{Kachru:2008yh,Taylor:2008tg} and \cite{Son:2008ye,Adams:2008wt}, from a pseudo-Riemannian coset construction by performing cosets on the Lifshitz and Schr\"odinger groups respectively.

However, aside from the somewhat counterintuitive concept of proposing locally relativistic spacetimes as duals to \nrel field theories \cite{Griffin:2012qx}, difficulties with field profile reconstruction \cite{Keeler:2014lia,Keeler:2013msa,Keeler:2015afa} as well as spacetime reconstruction \cite{Gentle:2015cfp} motivate the consideration of alternatives. As advocated in e.g. \cite{Hartong:2016yrf}, such \nrel spacetimes are more naturally viewed as Newton-Cartan spacetimes, described by a clock 1-form, degenerate spatial
metric and an extra $U(1)$ connection that contains the Newtonian potential. 
As we will see in this paper, the key observation is first of all that for certain choices of subgroup $\GroupH \subset \GroupG$ the invariant bilinear form on the coset is necessarily degenerate. 
Consequently there are two (degenerate)
distinct group-invariant bilinear forms on the coset: a covariant bilinear form ($\Omega_{ab}$) and a corresponding dual bilinear form ($\Omega^{ab}$). These objects are obviously not inverses of each other as they are both degenerate. With these two
objects, we can construct the Newton-Cartan geometric
data%
\footnote{See also Refs.~\cite{Brauner:2014jaa,Geracie:2015dea,Bekaert:2015xua,Karananas:2016hrm} for approaches
to Newton-Cartan geometry related to coset constructions, and \cite{Duval:2011mi,Duval:2016tzi} for other recent work on NC geometry.} for several interesting cases. 

Since some \nrel algebras can be obtained using  In\"on\"u-Wigner contractions of relativistic algebras, 
we also check our results from that perspective.  Moreover, as Newton-Cartan geometries can be obtained from a null reduction of a Riemmannian geometry in one dimension higher, we also obtain these embedding spaces.  In particular, we show that a judicious choice of $\GroupH$ results in a \ndeg bilinear form that gives rise to a pseudo-Riemannian spacetime with a null Killing vector via the standard coset construction. 

Even though we illustrate our ideas primarily with Newton-Cartan spaces, the coset procedure of this paper works in principle for more general cosets $\GroupG / \GroupH$, which can be non-reductive, can involve non-semisimple algebras, and can result in degenerate bilinear forms.  

We thus propose our generalized coset construction
as a generally applicable theory, which can also be used in other non-Lorentzian geometries. 

In Section \ref{sec:general}, we review both the standard coset construction and the Newton-Cartan geometry.  In Sections \ref{sec:bargmann}, \ref{sec:NH}, and \ref{sec:Schrodinger}, we develop coset spaces from the Bargmann, Newton-Hooke, and Schr\"odinger algebras respectively.  In Section \ref{sec:Sphere}, we present a Newton-Cartan analogue of the pseudo-Riemannian case of homogeneous spacetime with the topology of $\mathbb{R}\times S^2$. In fact, for the coset considered in that section, the bilinear form is not necessarily degenerate, but restricting to a degenerate form, it leads to this type of homogenous Newton-Cartan spacetime. Discussion of our results follows in \ref{sec:Discussion}. Again, while our focus is on Newton-Cartan, we also recognize that our coset procedure applies to other \nrel algebras as well. Therefore, the cosets for three other \nrel algebras, which do not produce Newton-Cartan geometry, are relegated to the appendices.


\section{Review of Cosets and Newton-Cartan Spaces}
\label{sec:general}

We now set our notation and review the coset construction for Riemannian or pseudo-Riemannian geometries.  We then review Newton-Cartan geometry and propose our generalization of the coset construction. Readers who are familiar with the standard coset construction may skip to the Newton-Cartan review in Section \ref{sec:NCcosetgeneral}.

\subsection{Coset construction for pseudo-Riemannian geometry}

Let $\Lieg$ be a finite-dimensional Lie algebra, $\Lieh \subset \Lieg$ be a subalgebra and $\cosetm = \Lieg \backslash \Lieh$ be the complement of $\Lieh$. Denote the elements of $\Lieg$ as
\begin{equation}
    \Lieg = \Span \{ \gel_{\gione} \}, \qquad \gione = 1, \ldots , | \Lieg |,
\end{equation}
where $| \Lieg |$ denotes the dimension of $\Lieg$. We split the $A$ index into two sets $\gione = ( \hione , \mione )$ with $\hione = 1 , \ldots, | \Lieh |$ labeling the generators in $\Lieh$ and $\mione = 0, \ldots , | \cosetm | -1$ labeling the generators in $\cosetm$. Without loss of generality, we assume that the basis $\gel_{\gione}$ for $\Lieg$ is suited to the split $\Lieg$ into $\cosetm$ and $\Lieh$, namely that the first $| \cosetm |$ are the generators in $\cosetm$ and the rest form the Lie subalgebra $\Lieh$. Therefore, we continue to use the symbol $\gel$ for generators in $\Lieh$ and $\cosetm$, with only the index to indicate which is which:
\begin{equation}
\begin{aligned}
    \Lieh &= \Span \{ \hel_{\hione} \}, \qquad \hione = 1, \ldots , | \Lieh |; \\
    \cosetm &= \Span \{ \mel_{\mione} \}, \qquad \mione = 0, \ldots , | \cosetm | -1.
\end{aligned}
\end{equation}
The numbering of the generators in $\cosetm$ goes from 0 to $|\cosetm | -1$ instead of 1 to $| \cosetm |$ because the local coordinates on the spacetime built out of $\cosetm$ will be dual to $\cosetm$ and so the numbering is chosen to coincide with the usual numbering of coordinates $x^{\mu}$, $\mu = 0, \ldots , | \cosetm | - 1$. We will often find it convenient to further split the $\mione$ indices into $\mione = 0$, which will usually correspond to the Hamiltonian, and $\mione = \miPone = 1, \ldots , | \cosetm | -1$, which will usually correspond to the momentum. Likewise, $\mu = 0$ ordinarily refers to the time coordinate and $\mu = i = 1, \ldots , | \cosetm | -1$ refers to spatial coordinates. Note that we use the same Latin index from the middle of the alphabet for both $\mione$-type and $\mu$-type indices.

We pass from algebra to group by exponentiation $\GroupG = \exp \Lieg$ and $\GroupH = \exp \Lieh$, associating each generator with its own dual coordinate. Then, we take the {\it left} coset $\cosetM = \GroupG / \GroupH$. Let $\Gel$ be an element of $\GroupG$, which we can write in terms of the dual coordinates as $\Gel=\exp{(x^\mu \delta_\mu^0 \mel_0)}\exp {(x^\mu \delta_\mu^1 \mel_1)}\ldots \exp{(x^\mu \delta_\mu^{|\cosetm|-1}\mel_{|\cosetm|-1})}$\footnote{In general any group element $\Gel$ that is parameterized in terms of $|m|$ coordinates that gives rise to nonzero invertible vielbeins $\viel^a$ on the coset space is an allowed coset representative.}. The Maurer-Cartan form associated with $\Gel$ is a $\Lieg$-valued one form, which we write as
\begin{equation}\label{eq:MCexplicit}
    \Gel^{-1} d \Gel = \mel_{\mione} \viel^{\mione} + \hel_{\hione} \gaugef^{\hione}= \mel_{\mione}\viel^{\mione}_\mu dx^\mu+\hel_{\hione} \gaugef^{\hione}_\mu dx^\mu,
\end{equation}
where $\viel^{\mione}$ is a vielbein on the coset space and $\gaugef^{\hione}$ are gauge fields associated with $\Lieh$, and in the last equality we have explicitly written the coordinate basis expression for clarity. The Maurer-Cartan form transforms under the adjoint representation, which thereby determines the general transformation rules of the fields.%
\footnote{This was done in~\cite{Hartong:2015zia} for the Bargmann algebra.} %
In this coset language, these transformations are simply the result of right-multiplying $\Gel$ by an element $\Hel \in \GroupH$, producing an equally good coset representative. The Maurer-Cartan form then transforms to
\begin{equation} \label{eq:MCformtrans}
    (\Gel \Hel )^{-1} d ( \Gel \Hel ) = \Hel^{-1} ( \Gel^{-1} d \Gel ) \Hel + \Hel^{-1} d \Hel.
\end{equation}
Once we have explicit expressions for $\Hel$ and $\Gel$, we may work out explicit expressions for the transformations of $\viel^{\mione}$ and $\gaugef^{\hione}$.
As will become relevant below, we note that usually one considers reductive cosets, meaning that the elements in $\cosetm$ transform in a representation of $\Lieh$. Our generalized coset procedure will involve choosing a subgroup $\GroupH$ in a way such that $\cosetm$ transforms in a projective representation of $\Lieh$.\footnote{Such choices of subgroup $\GroupH$ have been noted previously e.g. in \cite{Brauner:2014jaa}, but they were avoided for the reason that $\cosetm$ does not transform in a representation of $\Lieh$. However, it is only a mild complication that $\cosetm$ transforms under a projective representation instead.}

How do we construct the metric on the coset space? In the case when the coset space is a Riemannian manifold (of any signature), we first calculate the general solution for the symmetric bilinear form  $\Omega_{\mione \mitwo}$ on the coset space, satisfying the adjointness condition
\begin{equation} \label{eq:Omegalower}
    f_{\mione \hione}^{\hspace{0.3cm} \mithree} \Omega_{\mithree \mitwo}^{\phantom{c}} + f_{\mitwo \hione}^{\hspace{0.3cm} \mithree} \Omega_{\mithree \mione}^{\phantom{c}} = 0,
\end{equation}
where $f$ denotes the structure constants. This condition involves only components along $\cosetm$ of the commutators between an element of $\cosetm$ and an element of $\Lieh$. The metric is precisely $\Omega_{\mione \mitwo}$ when expressed in terms of the vielbein $\viel^{\mione}$:
\begin{equation} 
\label{eq:Omegalowermetric}
    ds^2 = \Omega_{\mione \mitwo} \viel^{\mione} \viel^{\mitwo}.
\end{equation}
For the Riemannian case, this metric must be \ndegafternoun, which forces $\Omega_{\mione \mitwo}$ to be \ndegafternoun as well.

There also exists a symmetric bilinear form with upper indices $\Omega^{\mione \mitwo}$ satisfying the adjointness condition
\begin{equation}
\label{eq:Omegaupper}
    \Omega^{\mione \mithree}f_{\mithree \hione}^{\hspace{0.3cm} \mitwo} + \Omega^{\mitwo \mithree} f_{\mithree \hione}^{\hspace{0.3cm} \mione} = 0.
\end{equation}
This generates its own metric, which is logically independent of \eqref{eq:Omegalowermetric}:
\begin{equation}
\label{eq:Omegauppermetric}
    \tilde{\partial}_{s}^{2} = \Omega^{\mione \mitwo} \viel_{\mione} \viel_{\mitwo},
\end{equation}
where $\viel_{\mione}$ is the inverse vielbein. In the Riemannian case, suitable coordinates can always be chosen to make $\Omega_{\mione \mitwo}$ and $\Omega^{\mione \mitwo}$ mutual inverses and then \eqref{eq:Omegauppermetric} is just the inverse metric to \eqref{eq:Omegalowermetric}.

As a simple example of this procedure, consider the Poincar\'e algebra for $\Lieg$ and the Lorentz subalgebra for $\Lieh$. Consider the coset representative $e^{\Pi_{\mione} \delta_{\mu}^{\mione} x^{\mu}}$, where $\Pi_{\mione}$ are the spacetime translation operators and $x^{\mu}$ are the spacetime coordinates. The Maurer-Cartan form is $\MCF = \Pi_{\mione} \delta_{\mu}^{\mione} dx^{\mu}$ and so the vielbein is flat, $\viel_{\mu}^{\mione} = \delta_{\mu}^{\mione}$. The general solution to (\ref{eq:Omegalower}), up to an overall prefactor, is $\Omega_{\mione \mitwo} = \eta_{\mione \mitwo}$. Therefore, the coset space is simply Minkowski space, $ds^2 = \Omega_{\mione \mitwo} \viel^{\mione} \viel^{\mitwo} = \eta_{\mu\nu} dx^{\mu} dx^{\nu}$. Deforming the Poincar\'e algebra by setting the commutator of two spacetime translations to be proportional to a Lorentz transformation (rather than zero) gives the de Sitter (dS) or anti-de Sitter (AdS) algebra, depending on the sign of the proportionality constant. The corresponding coset space is dS (or AdS) space. Another important example is the coset space representation of the $(n-1)$-sphere as the quotient of special orthogonal groups: $S^{n-1} \simeq SO(n)/SO(n-1)$.


\subsection{Newton-Cartan geometry}
\label{sec:NCcosetgeneral}

Since we will find that the Newton-Cartan geometry arises naturally out of our generalized coset construction, we now review some of its essential features. 

Instead of one \ndeg metric,
the metric data of NC geometry (also referred to below as the NC metric complex)  consist of a clock 1-form $\tau_\mu$,  inverse metric on space $h^{\mu\nu}$ and $U(1)$ gauge field $m_\mu$.\footnote{Often a connection is included in the description of Newton-Cartan geometry. While this is certainly necessary in order to construct a covariant derivative, it is a dependent object in that it is constructed from the information contained in $\tau$, $h$ and $m$. One convenient choice of metric compatible affine connections is obtained by demanding that the symmetric part of the Newton-Cartan connection is equal to the null reduced Levi-Civita connection of a higher-dimensional Lorentzian spacetime with a null isometry. This connection is also naturally suggested from the point of view of the Noether procedure applied to theories with Galilean symmetries \cite{Festuccia:2016awg}. We stress though that it remains a choice which metric compatible connection one uses. The physically important condition is that equations containing one or another connection are invariant under all Newton-Cartan gauge symmetries. For further discussions on the ambiguity among the class of Newton-Cartan metric compatible connections we refer to \cite{Bekaert:2014bwa, Hartong:2015xda, Bekaert:2015xua}}.

These objects admit the following local symmetries:
\begin{equation} \label{eq:TNCsym}
\begin{aligned}
\delta \tau_\mu &=  \mathcal{L}_\xi\tau_\mu\,,\qquad
\delta h^{\mu\nu}  =  \mathcal{L}_\xi h^{\mu\nu} \,, \\
\delta m_\mu &= \mathcal{L}_\xi m_\mu+\partial_\mu\sigma+\lambda_i \viel^i_\mu\,,
\end{aligned}
\end{equation}
where $h^{\mu\nu} = \delta^{ij} \viel_i^\mu \viel_j^\nu$. The vielbein and inverse vielbein satisfy the completeness relations $\viel_i^\mu \viel_\mu^j = \delta_i^j$ and  $-v^\mu\tau_\nu+ \viel^\mu_i \viel^i_\nu=\delta^\mu_\nu$, where $v^\mu$ is a velocity field satisfying $v^\mu \tau_\mu=-1$ and $v^\mu \viel_\mu^i=0$. The $\lambda_i$ parameterize local Galilean boosts (sometimes called Milne boosts), $\sigma$ parameterizes a $U(1)$ gauge transformation, and 
the Lie derivatives along $\xi^\mu$ correspond to the infinitesimal diffeomorphisms. The finite version of the infinitesimal diffeomorphism is just the standard transformation of a one-form under coordinate changes. The $U(1)$ transformation is the same whether infinitesimal or finite. The finite version of the Milne boost just picks up a quadratic piece in $\lambda_i$, which reads $\frac{1}{2} \lambda_i \lambda^i \tau$. The spatial vielbein also transforms under local tangent space spatial $SO(d)$ transformations. For use below we remark that one can construct the local Galilean boost invariant combination 
\begin{equation}
\label{eq:Phi} 
\tilde \Phi = -v^\mu m_\mu + \frac{1}{2} h^{\mu \nu} m_\mu m_\nu,
\end{equation}
which is related to the Newtonian potential.\footnote{An explicit expression for the boost invariant Newtonian potential appears in \cite{Duval:1993pe}; this can also be recovered directly from the expressions in \cite{Trumper:1983}, which first introduced the relevant and appropriate parametrization.  In this paper, we follow more closely the notation of \cite{Bergshoeff:2014uea}.} Depending on the properties of the clock 1-form $\tau_\mu$, one distinguishes
three cases (see e.g. \cite{Christensen:2013rfa}): (ordinary) NC geometry when  it is closed $d \tau =0$, twistless torsional NC geometry (TTNC) 
when it is hypersurface orthogonal $\tau \wedge d \tau =0$, and torsional
NC geometry (TNC) when $\tau_\mu$ is fully general.

NC geometry (and its torsionful generalization TNC geometry) can be obtained by gauging the Bargmann algebra \cite{Andringa:2010it,Hartong:2015zia} as follows. 
The Bargmann algebra consists of rotations $\qJ_{\miPone \miPtwo}$, Galilean boosts $\qG_{\miPone}$, the mass $\qN$, the Hamiltonian $\qH$ and momentum $\qP_{\miPone}$ with commutators
\begin{align} \label{eq:bargmannalgebra}
    [\qJ_{\miPone \miPtwo}, \qJ_{\miPthree \miPfour}] &= \delta_{\miPone \miPthree} \qJ_{\miPtwo \miPfour} - \delta_{\miPone \miPfour} \qJ_{\miPtwo \miPthree} - \delta_{\miPtwo \miPthree} \qJ_{\miPone \miPfour} + \delta_{\miPtwo \miPfour} \qJ_{\miPone \miPthree}, \notag \\
    [\qJ_{\miPone \miPtwo}, \qG_{\miPthree} ] &= \delta_{\miPone \miPthree} \qG_{\miPtwo} - \delta_{\miPtwo \miPthree} \qG_{\miPone}, \notag \\
    [\qJ_{\miPone \miPtwo}, \qP_{\miPthree} ] &= \delta_{\miPone \miPthree} \qP_{\miPtwo} - \delta_{\miPtwo \miPthree} \qP_{\miPone}, \\
    [\qH, \qG_{\miPone}] &= \qP_{\miPone}, \notag \\
    [\qP_{\miPone}, \qG_{\miPtwo} ] &= \qN \delta_{\miPone \miPtwo}. \notag
\end{align}
Here $\miPone, \miPtwo, \miPthree, \miPfour$ run over the spatial directions only.
One considers  the Bargmann-valued gauge field 
\begin{equation}
\mathcal{A}_\mu=H\tau_\mu+P_i \viel^i_\mu+Nm_\mu+\cdots,
\end{equation}
where we left out the connections associated with Galilean boosts $G_i$ and rotations $J_{ij}$.
The transformations \eqref{eq:TNCsym} then follow from the Bargmann algebra 
by considering the transformation $\delta\mathcal{A}_\mu=\mathcal{L}_\xi \mathcal{A}_\mu+\partial_\mu\Sigma+\left[\mathcal{A}_\mu,\Sigma\right]$, where $\xi^\mu$ generates diffeomorphisms and where $\Sigma=N\sigma+G_i\lambda^i+\frac{1}{2}J_{ij}\lambda^{ij}$ generates the local tangent space transformations. 

For the discussion below, we recall that these NC data and transformation properties can also be obtained
by null-reduction \cite{Duval:1984cj,Duval:1990hj,Julia:1994bs,Christensen:2013rfa}  from a Riemannian space with a 
 null Killing vector $\partial_u$.
The most general metric with this property is
\begin{equation}\label{eq:nullred}
G_{MN}dx^{M}dx^{N}=2\tau\left(du-m\right)+h_{\mu\nu}dx^\mu dx^\nu\,,
\end{equation}
where $\mu,\nu=0,1,...,d$, $M=(u,\mu)$ and $\tau=\tau_\mu dx^\mu$, $m=m_\mu dx^\mu$, $\text{det}\, h_{\mu\nu}=0$. The tensors $\tau_\mu$, $m_\mu$ and $h_{\mu\nu}=\delta_{ij} \viel_\mu^i \viel_\nu^j$  are independent of $u$. Note that from this point of view the $U(1)$ gauge transformation acts as $\delta u=\sigma$ and $\delta m=d\sigma$. 

The question we wish to address is how one recovers the NC metric complex for homogeneous spaces in a way analogous to the coset construction in the Riemannian case. It turns out that the coset construction follows through in logically the same manner. The main difference is that $\Omega_{\mione \mitwo}$ and $\Omega^{\mione \mitwo}$ will now be degenerate and therefore cannot be mutually inverse. The co- and contravariant $\Omega$ forms contain independent information. Interestingly, $\Omega_{\mione \mitwo}$ can be used to define the clock 1-form $\tau$, whereas $\Omega^{\mione \mitwo}$ will generate the (inverse) spatial metric $h^{\mu\nu}$. Furthermore we will often work with non-reductive cosets of the type that satisfy $[\Lieh,\cosetm]=\cosetm+\{N\}$ where $N$ is the generator associated with the NC connection $m$. There are cases where it is necessary to consider more general non-reductive cosets. An example will be given in section \ref{sec:Schrodinger} where we consider Schr\"odinger spacetimes. 

In the remainder of the paper, we will discuss our proposal in more detail, including how the NC  field $m_\mu$ enters, in four examples. We apply our proposal to two standard \nrel kinematical Lie algebras: the Bargmann and Newton-Hooke algebras. These cases have the added property of being In\"on\"u-Wigner contractions of relativistic algebras \cite{Bacry:1968zf}, a procedure which we show commutes with the coset construction. Thus, in these two cases, our proposed NC coset can be realized as a contraction of the standard (relativistic) coset procedure. We also offer examples using a non-kinematical algebra, namely the Schr\"odinger algebra, as well as examples of Newton-Cartan coset spaces with $SO(3)$ isometries. Finally, two smaller algebras (Aristotelian and Galilei) as well as the Carroll algebra are presented in the appendices. 
We expect that the construction is applicable more generally 
for cases with degenerate group-invariant bilinear forms.


\section{Newton-Cartan Spacetime from the Bargmann Group}
\label{sec:bargmann}

In this section we will consider cosets of the Bargmann group, 
whose algebra is given in \eqref{eq:bargmannalgebra} 
We will show how to construct flat Newton-Cartan spacetime as a coset of this group. Then, we will show two methods to derive this result from more standard scenarios, in which the $\Omega$-forms are \ndegafternoun: as a contraction of the relativistic algebra Poincar\'e$\, \oplus U(1)$ and as a null reduction of a pseudo-Riemannian spacetime in one higher dimension.


\subsection{Flat Newton-Cartan Spacetime}
\label{sec:Bargcoset}

We split the generators according to
\begin{align} \label{eq:Bargmannquotient}
    \Lieg &= \{ \qH, \qP_{\miPone} , \qJ_{\miPone \miPtwo}, \qG_{\miPone}, \qN \}, \notag \\
    \Lieh &= \{ \qJ_{\miPone \miPtwo}, \qG_{\miPone}, \qN \}, \\
    \qm &= \{ \qH, \qP_{\miPone} \}. \notag
\end{align}
The general solutions for the $\Omega$-forms in \eqref{eq:Omegalower} and \eqref{eq:Omegaupper} are
\begin{align} \label{eq:BargmannOmegas}
    \Omega_{\mione \mitwo} &= \Omega_{\qH \qH}
    \left( 
    \begin{array}{cc}
        1 & 0 \\
        0 & 0 
    \end{array}
    \right), &
    \Omega^{\mione \mitwo} &= \Omega^{\qP \qP}
    \left(
    \begin{array}{cc}
        0 & 0 \\
        0 & I
    \end{array}
    \right),
\end{align}
where $I$ is the identity matrix in the directions dual to the $\qP_\miPone$.

For the coset representative
\begin{equation} \label{eq:Bargmanncorep}
    \Gel = e^{\qH t} e^{\qP_{\miPone} x^{\miPone}},
\end{equation}
the Maurer-Cartan form is
\begin{equation} \label{eq:BargmannMC}
    \Gel^{-1} d \Gel = \qH dt + \qP_{\miPone} dx^{\miPone}.
\end{equation}
As in the \ndeg case \eqref{eq:MCexplicit}, we read off the vielbeine from the coefficient of the generators in $\qm$, while the gauge fields $m^\hione$ come from the coefficients of the generators in $\Lieh$.
We see that 
$m=0$ and the vielbeine are $\viel^{\mione} = ( \tau , \viel^{\miPone} )$, where
\begin{align} \label{eq:FNCtauviel}
    \tau &= dt, &
    \viel^{\miPone} &= dx^\mu\delta_\mu^{\miPone}=dx^\miPone,
\end{align}
where we slightly abuse notation in the last equality.\footnote{There is no harm as long as we remember that the vielbein $\viel^{\mione}$ is a one form in the coset space coordinates $x^\mu$, with group index matching the generator $\qP_\miPone$.  Here (and often) it just turns out that the vielbein $\viel^\mione$ only has legs in the coordinate direction $\mu=\mione$ dual to the same generator $\qP_\miPone$.
}
Therefore, after scaling coordinates to set $\Omega_{\qH \qH} = \Omega^{\qP \qP} = 1$, the parts of the metric complex coming from $\Omega_{\qA \qB}$ and $\Omega^{\qA \qB}$ are
\begin{equation}
\begin{aligned}
    \Omega_{\mione \mitwo} \viel^{\mione} \viel^{\mitwo} &= \tau^2 = dt^2, \\
    \Omega^{\mione \mitwo} \viel_{\mione} \viel_{\mitwo} &= \viel_{\mione} \viel_{\mione} = h^{\mu\nu} \partial_{\mu} \partial_{\nu} = \partial_{\miPone} \partial_{\miPone}.
\end{aligned}
\end{equation}
Thus the result of this coset construction is:
\begin{align}  \label{eq:flatNCmetriccomplex}
    \tau &= dt, \notag \\
    h^{\mu\nu} &= \delta_{\miPone}^{\mu} \delta_{\miPone}^{\nu}, \\
    m &= 0. \notag
\end{align}
which is flat Newton-Cartan spacetime. This geometry is torsion-free Newton-Cartan (NC) since $d \tau = 0$.

Suppose we right multiply (\ref{eq:Bargmanncorep}) by $\Hel = e^{\qG_{\miPone} \lgb^{\miPone}} e^{\qN \lut}$, which generates a local Galilean boost and $U(1)$ transformation. Using the transformation of the Maurer-Cartan form (\ref{eq:MCformtrans}), we can determine the transformation of the Newton-Cartan data,
\begin{align} \label{eq:tauemtrans}
    \tau &\rightarrow \tau' = \tau, \notag \\
    \viel^{\miPone} &\rightarrow \viel^{\miPone}{}' = \viel^{\miPone} + \lgb^{\miPone} \tau, \\
    m &\rightarrow m' = m + \lgb_{\miPone} \viel^{\miPone} + \frac{1}{2} \lgb_{\miPone} \lgb^{\miPone} \tau + d \lut. \notag
\end{align}
As expected, this precisely reproduces the known transformation properties of the Newton-Cartan metric complex under local Galilean boosts and $U(1)$ transformations.


\subsection{Bargmann as Contracted Poincar\'e\texorpdfstring{$\, \oplus U(1)$}{xU(1)}}
\label{sec:Bargcontr}

The coset construction described previously, which naturally produces Newton-Cartan geometry, may seem exotic and the appearance of degenerate bilinear forms $\Omega$ may be unsettling. Indeed, the existence of degenerate $\Omega$ has been noted in the past (in \cite{SchaferNameki:2009xr} and \cite{Brauner:2014jaa}, for instance), but avoided, perhaps because the connection to Newton-Cartan had not been appreciated or desired.

For this reason, we allay these concerns by showing that the degenerate forms and the Newton-Cartan metric complex in the previous Bargmann case arise as an In\"on\"u-Wigner contraction of a perfectly ordinary relativistic coset space.\footnote{See, e.g., \cite{Ballesteros:1994, Herranz:2002} for previous work on the geometry of nonrelativistic kinematical groups where the issue of degenerate metrics as well as their contraction theory is addressed.} The relativistic algebra in this case is the Poincar\'e algebra with an additional $U(1)$. The Poincar\'e generators are Lorentz generators $M_{\mione \mitwo}$ and spacetime translations $\Pi_{\mione}$ with commutators in $(-, +, \ldots , +)$ signature,
\begin{equation} \label{eq:Poincare}
\begin{aligned} 
    \phantom{i} [ \qrJ_{\mione \mitwo}, \qrJ_{\mithree \mifour} ] &= \eta_{\mione \mithree} \qrJ_{\mitwo \mifour} - \eta_{\mione \mifour} \qrJ_{\mitwo \mithree} - \eta_{\mitwo \mithree} \qrJ_{\mione \mifour} + \eta_{\mitwo \mifour} \qrJ_{\mione \mithree}, \\
    [\qrJ_{\mione \mitwo} , \qrP_{\mithree}] &= \eta_{\mione \mithree} \qrP_{\mitwo} - \eta_{\mitwo \mithree} \qrP_{\mione}.
\end{aligned}
\end{equation}
The $U(1)$ generator is $Z$ and commutes with everything. We now introduce a constant $c$ (speed of light) and change to a new basis of generators 
\begin{equation} \label{eq:Bargcontract}
\begin{aligned}
    \qJ_{\miPone \miPtwo} &= \qrJ_{\miPone \miPtwo}, \\
    \qG_{\miPone} &= \frac{1}{\qc} \qrJ_{0 \miPone}, \\
    \qN &= \frac{1}{2 \qc} ( \qrP_0 - Z ), \\
    \qH &= \qc ( \qrP_0 + Z ), \\
    \qP_{\miPone} &= \qrP_{\miPone}.
\end{aligned}
\end{equation}
The commutators in this basis read
\begin{align} \label{eq:Bargcontractalg}
    [\qJ_{\miPone \miPtwo}, \qJ_{\miPthree \miPfour}] &= \delta_{\miPone \miPthree} \qJ_{\miPtwo \miPfour} - \delta_{\miPone \miPfour} \qJ_{\miPtwo \miPthree} - \delta_{\miPtwo \miPthree} \qJ_{\miPone \miPfour} + \delta_{\miPtwo \miPfour} \qJ_{\miPone \miPthree}, \notag \\
    [\qJ_{\miPone \miPtwo}, \qG_{\miPthree} ] &= \delta_{\miPone \miPthree} \qG_{\miPtwo} - \delta_{\miPtwo \miPthree} \qG_{\miPone}, \notag \\
    [\qJ_{\miPone \miPtwo}, \qP_{\miPthree} ] &= \delta_{\miPone \miPthree} \qP_{\miPtwo} - \delta_{\miPtwo \miPthree} \qP_{\miPone}, \notag \\
    [\qH, \qG_{\miPone}] &= \qP_{\miPone}, \\
    [\qP_{\miPone}, \qG_{\miPtwo} ] &= \biggl( \qN + \frac{1}{2\qc^2} \qH \biggr) \delta_{\miPone \miPtwo}, \notag \\
    [ \qG_{\miPone} , \qG_{\miPtwo} ] &= - \frac{1}{\qc^2} \qJ_{\miPone \miPtwo}, \notag \\
    [ \qN , \qG_{\miPone} ] &= \frac{1}{2 \qc^2} \qP_{\miPone}. \notag
\end{align}
The Bargmann algebra (\ref{eq:bargmannalgebra}) is then  recovered by the contraction limit $\qc \rightarrow \infty$.

We now take the same split of generators as in (\ref{eq:BargmannOmegas}) and use the algebra \eqref{eq:Bargcontractalg}. The resulting $\Omega$-forms are
\begin{align} \label{eq:BargOsb4contract}
    \Omega_{\mione \mitwo} &= \Omega_{\qH \qH}
    \left( 
    \begin{array}{cc}
        1 & 0 \\
        0 & - \frac{1}{2 \qc^2} I 
    \end{array}
    \right), &
    \Omega^{\mione \mitwo} &= \Omega^{\qP \qP}
    \left(
    \begin{array}{cc}
        - \frac{1}{2 \qc^2} & 0 \\
        0 & I
    \end{array}
    \right),
\end{align}
which reproduce the Bargmann result (\ref{eq:BargmannOmegas}) in the $\qc \rightarrow \infty$ limit. Thus, the degenerate $\Omega$-forms for the case of Bargmann can be understood as the singular limit of the \ndeg relativistic case above.

Since $\qH$ and $\qP_{\miPone}$ still commute, we take the same coset representative (\ref{eq:Bargmanncorep}) and find the same Maurer-Cartan form, $m=0$ and flat Newton-Cartan metric complex (\ref{eq:FNCtauviel}).
Thus we have established that our generalized coset construction is consistent with the \nrel limit of the Poincar\'e coset. 


\subsection{Flat Newton-Cartan via Null Reduction}
\label{sec:flatNCnullreduction}

In \eqref{eq:Bargmannquotient} we chose to put the mass generator $\qN$ of the Bargmann algebra in $\Lieh$. We now examine what happens when we instead place  $\qN$ in $\cosetm$: 
\begin{align} \label{eq:BargmannquotientNR}
    \Lieg &= \{ \qN, \qH, \qP_{\miPone} , \qJ_{\miPone \miPtwo}, \qG_{\miPone} \}, \notag \\
    \Lieh &= \{ \qJ_{\miPone \miPtwo}, \qG_{\miPone} \}, \\
    \qm &= \{ \qN , \qH, \qP_{\miPone} \}. \notag
\end{align}
Using the algebra in \eqref{eq:Bargcontractalg}, the general solutions to \eqref{eq:Omegalower}, \eqref{eq:Omegaupper} for the $\Omega$-forms are
\begin{equation} \label{eq:BargmannNROs}
\begin{aligned}
    \Omega_{\mione \mitwo} &= 
    \left( 
    \begin{array}{ccc}
        \frac{1}{4c^4} \Omega_{\qH \qH} & - \Omega_{\qP \qP} - \frac{1}{2c^2} \Omega_{\qH \qH} & 0 \\
        - \Omega_{\qP \qP} - \frac{1}{2c^2} \Omega_{\qH \qH} & \Omega_{\qH \qH} & 0 \\
        0 & 0 & \Omega_{\qP \qP} I
    \end{array}
    \right), \\
    \Omega^{\mione \mitwo} &=
    \left(
    \begin{array}{ccc}
        \Omega^{\qN \qN} & - \Omega^{PP} - \frac{1}{2c^2} \Omega^{\qN \qN} & 0 \\
        - \Omega^{\qP \qP} - \frac{1}{2c^2} \Omega^{\qN \qN} & \frac{1}{4c^4} \Omega^{\qN \qN} & 0 \\
        0 & 0 & \Omega^{\qP \qP} I
    \end{array}
    \right),
\end{aligned}
\end{equation}
where $I$ is the identity matrix in the $P$ directions. As expected for a relativistic coset, these forms are \ndegafternoun and, in fact, they can be made mutually inverse by
\begin{align} \label{eq:BargNRinvrels}
    \Omega^{\qP \qP} &= \frac{1}{\Omega_{\qP \qP}}, &%
    \Omega^{\qN \qN} &= - \frac{c^2 \Omega_{\qH \qH}}{\Omega_{\qP \qP} \bigl( \Omega_{\qH \qH} + c^2 \Omega_{\qP \qP} \bigr)}.
\end{align}

We choose the coset representative to be
\begin{equation} \label{eq:Bargcontrcorep}
    \Gel = e^{\qH t} e^{\qP_{\miPone} x^{\miPone}} e^{- \qN u}.
\end{equation}
Note the minus sign in the exponential for $\qN$, which is only there to conform with the conventions of the null reduction.

The Maurer-Cartan form is flat,
\begin{equation} \label{eq:MCbargnullred}
    \Gel^{-1} d \Gel = \qH dt + \qP_{\miPone} dx^{\miPone} - \qN du.
\end{equation}
In order to perform a null reduction, we require a geometry with a null Killing vector, which is usually associated with the $u$ coordinate. Therefore, we choose to take $\Omega_{\qH \qH} = 0$ (and, correspondingly, $\Omega^{\qN \qN} = 0$), and we scale coordinates so as to set $\Omega_{\qP \qP} = \Omega^{\qP \qP} = 1$. With this choice, the metric reads
\begin{equation} \label{eq:BargNRmet}
    ds^2 = 2dt \, du + \delta_{\miPone \miPtwo} dx^{\miPone} dx^{\miPtwo}.
\end{equation}
This metric is written in the form suitable for null reduction and, comparing to \eqref{eq:nullred}, we find that the Newton-Cartan metric complex is precisely the one we found earlier, in \eqref{eq:flatNCmetriccomplex}. 

\section{Newton-Cartan Spacetime from the Newton-Hooke Group}
\label{sec:NH}

We now turn to a case that corresponds to a more non-trivial NC spacetime. To this end we consider cosets of the Newton-Hooke group.  The Newton-Hooke algebra is a deformation of the Bargmann algebra by one extra commutator parametrized by a cosmological constant $\Lambda$, given by 
\begin{equation} \label{eq:NHalgebra}
\begin{aligned}
    \phantom{i} [\qJ_{\miPone \miPtwo}, \qJ_{\miPthree \miPfour}] &= \delta_{\miPone \miPthree} \qJ_{\miPtwo \miPfour} - \delta_{\miPone \miPfour} \qJ_{\miPtwo \miPthree} - \delta_{\miPtwo \miPthree} \qJ_{\miPone \miPfour} + \delta_{\miPtwo \miPfour} \qJ_{\miPone \miPthree}, \\
    [\qJ_{\miPone \miPtwo}, \qG_{\miPthree} ] &= \delta_{\miPone \miPthree} \qG_{\miPtwo} - \delta_{\miPtwo \miPthree} \qG_{\miPone}, \\
    [\qJ_{\miPone \miPtwo}, \qP_{\miPthree} ] &= \delta_{\miPone \miPthree} \qP_{\miPtwo} - \delta_{\miPtwo \miPthree} \qP_{\miPone}, \\
    [\qH, \qG_{\miPone}] &= \qP_{\miPone}, \\
    [\qP_{\miPone}, \qG_{\miPtwo} ] &= \qN \delta_{\miPone \miPtwo} \\
    [\qH , \qP_{\miPone}] &= \Lambda \qG_{\miPone}.
\end{aligned}
\end{equation}
In the sequel, we will assume $\Lambda > 0$ and show that the coset space can be put in a form that resembles, but is distinct from, de Sitter space. Alternatively, it can be interpreted as flat Newton-Cartan spacetime with a quadratic potential $m$ field. The $\Lambda < 0$ case follows mutatis mutandis. 
We will refer to this particular NC spacetime as
Newton-Hooke spacetime.


\subsection{Newton-Hooke Spacetime}
\label{sec:NHcoset}

As in the first Bargmann coset in Sec. \ref{sec:Bargcoset}, we split the generators according to
\begin{align} \label{eq:NHquotient}
    \Lieg &= \{ \qH, \qP_{\miPone} , \qJ_{\miPone \miPtwo}, \qG_{\miPone}, \qN \}, \notag \\
    \Lieh &= \{ \qJ_{\miPone \miPtwo}, \qG_{\miPone}, \qN \}, \\
    \qm &= \{ \qH, \qP_{\miPone} \}. \notag
\end{align}
The general solutions for the $\Omega$-forms are the same as (\ref{eq:BargmannOmegas}) in the Bargmann case
\begin{align} \label{eq:NHOmegas}
    \Omega_{\mione \mitwo} &= \Omega_{\qH \qH}
    \left( 
    \begin{array}{cc}
        1 & 0 \\
        0 & 0 
    \end{array}
    \right), &
    \Omega^{\mione \mitwo} &= \Omega^{\qP \qP}
    \left(
    \begin{array}{cc}
        0 & 0 \\
        0 & I
    \end{array}
    \right),
\end{align}
Consider the same parametrization of coset representative as (\ref{eq:Bargmanncorep}) in the Bargmann case,
\begin{equation} \label{eq:NHcorep}
    \Gel = e^{\qH t} e^{\qP_{\miPone} x^{\miPone}}.
\end{equation}
Since $\qH$ and $\qP_{\miPone}$ no longer commute, the Maurer-Cartan form is more complicated now:
\begin{equation} \label{eq:NHMC1}
    \Gel^{-1} d \Gel = \qH dt + \qP_{\miPone} dx^{\miPone} - \frac{1}{2} \Lambda \qN x^2 dt + \Lambda \qG_{\miPone} x^{\miPone} dt,
\end{equation}
where $x^2 = x_{\miPone} x^{\miPone}$. The vielbeine and gauge fields become
\begin{align}
\tau &= dt\\
\viel^\miPone &= dx^\miPone\\
m^N \equiv m &= -\frac{1}{2}\Lambda x^2 dt.
\end{align}
Note there is another non-zero field $m^{G_\miPone}=\Lambda x^\miPone dt$. However, this is  
 not an independent connection, since
the Galilean boost and rotation connections for NC geometry
can be written in terms of the NC geometric data $\tau, \viel^{\miPone}, m$ \cite{Andringa:2010it}. 
Contracting with the $\Omega$, we find flat Newton-Cartan geometry with a quadratic Newton potential,
\begin{align} \label{eq:NHcomp}
    \tau &= dt, \notag \\
    h^{\mu\nu} &= \delta_{i}^{\mu} \delta_{i}^{\nu}, \\
    m &= - \frac{1}{2} \Lambda x^2 dt. \notag
\end{align}
where we recall that the Newton potential is related to
$m$ via \eqref{eq:Phi}. This geometry is also torsionless NC since $d \tau = 0$.

As pointed out in \cite{Bekaert:2014bwa}, for NC geometry
(zero torsion) one can always use local Galilean boosts and $U(1)$ transformations to remove $m$ at the cost of introducing potentially complicated coordinates. The clock form, vielbein and $m$ field transform as in (\ref{eq:tauemtrans}). We seek $\lgb^{\miPone}$ and $\lut$ such that
\begin{align}
    0 = m' &= m + \lgb_{\miPone} \viel^{\miPone} + \frac{1}{2} \lgb_{\miPone} \lgb^{\miPone} \tau + d \lut \notag \\
    &= - \frac{1}{2} \Lambda x^2 dt + \lgb_{\miPone} dx^{\miPone} + \frac{1}{2} \lgb_{\miPone} \lgb^{\miPone} dt + d \lut.
\end{align}
The general solution for $\lgb^{\miPone}$ and $\lut$ can be parametrized as
\begin{equation} \label{eq:lgblut}
\begin{aligned}
    \lgb^{\miPone} &= \sqrt{\Lambda} \lgbm (t) x^{\miPone}, \\
    \lut &= - \frac{1}{2} \sqrt{\Lambda} \lgbm (t) x^2,
\end{aligned}
\end{equation}
where the function $\lgbm (t)$ satisfies
\begin{equation}
    \dot{\lgbm} = \sqrt{\Lambda} ( \lgbm^2 -1).
\end{equation}
The general solution for $\lgbm (t)$ is
\begin{equation}
    \lgbm (t) = - \tanh \left[ \sqrt{\Lambda} (t-t_0) \right],
\end{equation}
where $t_0$ is some constant.

The transformed vielbein is given by
\begin{equation}
    \viel^{\miPone}{}' = \viel^{\miPone} + \lgb^{\miPone} \tau = dx^{\miPone} + \sqrt{\Lambda} \lgbm (t) x^{\miPone} dt = \frac{1}{\lgbe (t)} d \left[ \lgbe (t) x^{\miPone} \right],
\end{equation}
where the function $\lgbe (t)$ satisfies
\begin{equation}
    \dot{\lgbe} = \sqrt{\Lambda} \lgbm \lgbe.
\end{equation}
The general solution for $\lgbe (t)$ is
\begin{equation}
    \lgbe (t) = C \, \sech \left[ \sqrt{\Lambda} (t-t_0) \right],
\end{equation}
where $C$ is some constant.

Therefore, we define the new set of coordinates
\begin{equation} \label{eq:coordtrans}
\begin{aligned}
    t' &= t, \\
    x^{\miPone}{}' &= \lgbe (t) x^{\miPone},
\end{aligned}
\end{equation}
in terms of which, the new Newton-Cartan data read
\begin{align}
    \tau' &= dt', \notag \\
    h^{\mu\nu}{}' &= \lgbe^2 ( t' ) \, \delta_{i}^{\mu} \delta_{i}^{\nu}, \\
    m' &= 0. \notag
\end{align}
There are some interesting specific choices one can make for $\lgbm (t)$ and $\lgbe (t)$. For example, set $C = e^{- \sqrt{\Lambda} t_0}$ and take the limit $t_0 \rightarrow - \infty$, in which case $\lgbm (t) = -1$ and $\lgbe (t) = e^{- \sqrt{\Lambda} t}$. Then, the spatial metric resembles the flat slicing of de Sitter space,
\begin{equation} \label{eq:NHdeSitterlike}
    h_{\mu\nu} dx^{\mu} dx^{\nu} = \frac{1}{\lgbe^2 (t)} dx_{\miPone} dx^{\miPone} = e^{2 \sqrt{\Lambda} t} dx_{\miPone} dx^{\miPone},
\end{equation}
where we drop the primes since there is no longer any need to refer to the original coordinates that gave (\ref{eq:NHcomp}). Note that this spatial metric plus the clock 1-form $\tau=dt$ build a metric complex, not a metric. Importantly, the space described by this metric complex is not de Sitter; it lacks local Lorentz symmetry and thus has the wrong tangent space to be de Sitter.

Another limit we can take is to set $C=1$ and $t_0 = 0$. In this case,
\begin{equation} \label{eq:coshmetric}
    h_{\mu\nu} dx^{\mu} dx^{\nu} = \cosh^2 \left( \sqrt{\Lambda} t \right) dx_{\miPone} dx^{\miPone}.
\end{equation}
In fact, one could have derived this latter limit directly just at the level of the choice of coset representative. Note that one obvious way to force $m=0$ is to switch the order of $\qH$ and $\qP_{\miPone}$ in (\ref{eq:NHcorep}):
\begin{equation}
    \Gel = e^{\qP_{\miPone} x^{\miPone}} e^{\qH t},
\end{equation}
in which case, the Maurer-Cartan form reads
\begin{equation} \label{eq:NHMC2}
    \Gel^{-1} d \Gel = \qH dt + \qP_i \cosh \left( \sqrt{\Lambda} t \right) dx^i - \sqrt{\Lambda} \qG_i \sinh \left( \sqrt{\Lambda} t \right) dx^i.
\end{equation}
Indeed, this reproduces (\ref{eq:coshmetric}) along with $\tau = dt$ and $m=0$. 

This has the somewhat surprising corollary that
\begin{equation} \label{eq:NHid}
    e^{\qP_{\miPone} x^{\miPone}{}'} e^{\qH t} = e^{\qH t} e^{\qP_{\miPone} x^{\miPone}} e^{\qG_{\miPone} \lgb^{\miPone}} e^{\qN \lut},
\end{equation}
where $\lgb^{\miPone}$ and $\lut$ are given by (\ref{eq:lgblut}) and $x^{\miPone}{}'$ is given by (\ref{eq:coordtrans}) with $\lgbm (t) = - \tanh \bigl( \sqrt{\Lambda} t \bigr)$ and $\lgbe (t) = \sech \bigl( \sqrt{\Lambda} t \bigr)$. One can prove (\ref{eq:NHid}) directly as follows. First,
\begin{align}
    e^{-\qH t} e^{\qP_{\miPone} x^{\miPone}{}'} e^{\qH t} &= \exp \left( e^{- \qH t} \qP_{\miPone} x^{\miPone}{}' e^{\qH t} \right) \notag \\
    &= \exp \left[ \cosh \left( \sqrt{\Lambda} t \right) \qP_{\miPone} x^{\miPone}{}' - \sqrt{\Lambda} \sinh \left( \sqrt{\Lambda} t \right) \qG_{\miPone} x^{\miPone}{}' \right] \notag \\
    &= e^{\qP_{\miPone} x^{\miPone} + \qG_{\miPone} \lgb^{\miPone}},
\end{align}
where use has been made of the relation between $x^{\miPone}{}'$ and $x^{\miPone}$ and the form of $\lgb^{\miPone}$. Now, we may use the Zassenhaus formula,
\begin{align} 
    e^{\qP_{\miPone} x^{\miPone} + \qG_{\miPone} \lgb^{\miPone}} &= e^{\qP_{\miPone} x^{\miPone}} e^{\qG_{\miPone} \lgb^{\miPone}} e^{- \frac{1}{2} x^{\miPone} \lgb^{\miPtwo} [ \qP_{\miPone} , \qG_{\miPtwo} ]} \cdots \notag \\
    &= e^{\qP_{\miPone} x^{\miPone}} e^{\qG_{\miPone} \lgb^{\miPone}} e^{- \frac{1}{2} \qN \lgb_{\miPone} x^{\miPone}} \notag \\
    &= e^{\qP_{\miPone} x^{\miPone}} e^{\qG_{\miPone} \lgb^{\miPone}} e^{\qN \lut},
\end{align}
where $\cdots$ denotes products of exponentials of nested commutators of $\qP_{\miPone}$ and $\qG_{\miPone}$, which all vanish since the first commutator is proportional to $\qN$. We have used the identity $\sigma = - \frac{1}{2} \lgb_{\miPone} x^{\miPone}$, which is true regardless of the specific choices of $\lgbm (t)$ and $\lgbe (t)$.


\subsection{Newton-Hooke as Contracted dS/AdS\texorpdfstring{$\, \oplus U(1)$}{xU(1)}}
\label{sec:NHcontr}

The Newton-Hooke result can be derived using an In\"on\"u-Wigner contraction of a relativistic algebra in much the same way as the Bargmann algebra in Sec. \ref{sec:Bargcontr}. 

We deform the Poincar\'e algebra (\ref{eq:Poincare}) with the commutator
\begin{equation} \label{eq:PiPi}
    [\qrP_{\mione}, \qrP_{\mitwo}] = \lambda \qrJ_{\mione \mitwo},
\end{equation}
where $\lambda$ is some constant. We call this algebra the de Sitter (Anti-de Sitter) algebra for positive (negative) $\lambda$.

With the identifications for $\qJ_{\miPone \miPtwo}$, $\qG_{\miPone}$, $\qN$, $\qH$ and $\qP_{\miPone}$ given in (\ref{eq:Bargcontract}), only three additional commutators arise in addition to (\ref{eq:Bargcontractalg}),
\begin{equation} \label{eq:NHcontractalg}
\begin{aligned}[]
    [H,P_i] &= \Lambda G_i, \\
    [P_i, P_j] &= \frac{\Lambda}{\qc^2} J_{ij}, \\
    [N,P_i] &= \frac{\Lambda}{2 \qc^2} G_i,
\end{aligned}
\end{equation}
where we have introduced the cosmological constant factor
\begin{equation} \label{eq:Lambda}
    \Lambda = \qc^2 \lambda.
\end{equation}
Indeed, the $c \rightarrow \infty$ limit exactly reproduces the Newton-Hooke algebra \eqref{eq:NHalgebra}. 

We parametrize a coset representative as
\begin{equation} \label{eq:gNH}
    g = e^{Ht} e^{P_d x^d} e^{P_{d-1} x^{d-1}} \cdots e^{P_1 x^1}.
\end{equation}
Note that the order of the $P$ terms actually matters now because they no longer commute with each other! Here, $d$ is the number of spatial dimensions. We find the vielbeine and $m$ gauge field,
\begin{align} \label{eq:NHcontrcomps}
    \tau &= \frac{1}{2} dt \left[ 1 + \prod_{\miPtwo =1}^{d} \cos \left( \frac{\sqrt{\Lambda} x^{\miPtwo}}{\qc} \right) \right], \notag \\
    \viel^{\miPone} &= dx^{\miPone} \prod_{\miPtwo < \miPone} \cos \left( \frac{\sqrt{\Lambda} x^{\miPtwo}}{\qc} \right), \\
    m^{\qN} &\equiv m = - \qc^2 dt \left[ 1 - \prod_{\miPtwo =1}^{d} \cos \left( \frac{\sqrt{\Lambda} x^{\miPtwo}}{\qc} \right) \right]. \notag
\end{align}
The $\Omega$-forms are the same as in the Bargmann case in \eqref{eq:BargOsb4contract}. The results of the Newton-Hooke coset are reproduced in the $c \rightarrow \infty$ limit. For example, in this limit, $m$ is given by
\begin{equation} \label{eq:mlim}
    m = - \qc^2 dt \left[ 1 - \prod_{j=1}^{d} \left( 1 - \frac{\Lambda (x^j)^2}{2 \qc^2} + O(\qc^{-4}) \right) \right] = - \frac{1}{2} \Lambda x^2 dt + O(\qc^{-2}),
\end{equation}
where $x^2 = x_i x^i$.


\subsection{Newton-Hooke Spacetime via Null Reduction}
\label{sec:NHNR}

As in the Bargmann case, we can also derive the previous Newton-Hooke coset result via null reduction by placing $\qN \in \cosetm$. Note that the additional commutators in \eqref{eq:NHcontractalg}, compared to the Bargmann case, involve only commutators of elements which are now all in $\cosetm$. Therefore, the relevant structure constants in the relations \eqref{eq:Omegalower} and \eqref{eq:Omegaupper} defining the $\Omega$ forms are unchanged. Likewise, the general solutions for the $\Omega$ forms are the same as (\ref{eq:BargmannNROs}) in the Bargmann case.

The coset representative is chosen to be the same as (\ref{eq:Bargcontrcorep}) in the Bargmann case. The Maurer-Cartan form is
\begin{equation} \label{eq:NHcontrMC}
    \Gel^{-1} d \Gel = \qH dt + \qP_{\miPone} dx^{\miPone} - \qN \biggl( du + \frac{1}{2} \Lambda x^2 dt \biggr) + \Lambda ( \qG_{\miPone} x^{\miPone} ) dt.
\end{equation}
Again, with the choice $\Omega_{\qH \qH} = \Omega^{\qN \qN} = 0$ and $\Omega_{\qP \qP} = \Omega^{\qP \qP} = 1$, the metric is
\begin{equation} \label{eq:NHNRmet}
    ds^2 = 2dt \biggl( du + \frac{1}{2} \Lambda x^2 \, dt \biggr) + \delta_{\miPone \miPtwo} dx^{\miPone} dx^{\miPtwo}.
\end{equation}
Comparing with \eqref{eq:nullred}, this exactly reproduces the metric complex in \eqref{eq:NHcomp}.\footnote{See \cite{Gibbons:2003rv} for previous work on the coset construction of Newton-Hooke spacetimes and their description via null reduction. Note that the algebra used therein to construct the Newton-Hooke spacetime does not contain the central element $N$ and therefore the construction does not include the gauge field $m_{\mu}$. However, when the spacetime is derived via null reduction, this central element is indeed included.}


\section{Newton-Cartan Spacetime from the Schr\"odinger Group}
\label{sec:Schrodinger}

We now examine cosets based on the Schr\"odinger algebra, which is the extension of the Bargmann algebra 
by a dilatation operator $\qD$. The non-zero commutators added to \ref{eq:bargmannalgebra} are
\begin{equation} \label{eq:SchDalgebra}
\begin{aligned}
    \phantom{i} [ \qD , \qH ] &= -z \qH, \\
    [ \qD , \qP_{\miPone} ] &= - \qP_{\miPone}, \\
    [ \qD , \qG_{\miPone} ] &= (z-1) \qG_{\miPone}, \\
    [ \qD , \qN ] &= (z-2) \qN,
\end{aligned}
\end{equation}
where $z$ is the dynamical critical exponent.

We will focus on the case of $2+1$ dimensions and $z=2$, which is special in that one can add a generator $\qK$ for special conformal transformations as well:
\begin{equation} \label{eq:SchDKalgebra}
\begin{aligned}
    \phantom{i} [ \qJ , \qG_{\miPone} ] &= \epsilon_{\miPone}^{\phantom{\miPone} \miPtwo} \qG_{\miPtwo}, &\qquad%
    [ \qJ , \qP_{\miPone} ] &= \epsilon_{\miPone}^{\phantom{\miPone} \miPtwo} \qP_{\miPtwo}, \\
    [ \qH , \qG_{\miPone} ] &= \qP_{\miPone}, &\qquad%
    [ \qP_{\miPone} , \qG_{\miPtwo} ] &= \qN \delta_{\miPone \miPtwo}, \\
    [ \qD , \qH ] &= -2 \qH, &\qquad%
    [ \qD , \qP_{\miPone} ] &= - \qP_{\miPone}, \\
    [ \qD , \qG_{\miPone} ] &= \qG_{\miPone}, &\qquad%
    [ \qD , \qK ] &= 2 \qK, \\
    [ \qK , \qH ] &= - \qD, &\qquad%
    [ \qK , \qP_{\miPone} ] &= - \qG_{\miPone}.
\end{aligned}
\end{equation}
Here $J = J_{12}$ and $\epsilon$ is the antisymmetric symbol with sign convention $\epsilon_{1}^{\phantom{1} 2}$ = 1.

This algebra is not the contraction of any relativistic algebra, and so we cannot derive the coset via a contraction as we did for the Bargmann and Newton-Hooke algebras. However, now that we are confident in how the coset construction works in the Newton-Cartan case, we will soldier on and demonstrate that the coset produces a Newton-Cartan spacetime with $z=2$ Lifshitz scaling.


\subsection{Newton-Cartan Spacetime with Lifshitz Scaling Manifest}
\label{sec:Schrodcoset}

We split the generators according to
\begin{align} \label{eq:Schrodquotient}
    \Lieg &= \{ \qH, \qP_{\miPone} , \qD , \qJ, \qG_{\miPone}, \qN, \qK \}, \notag \\
    \Lieh &= \{ \qJ, \qG_{\miPone}, \qN, \qK \}, \\
    \qm &= \{ \qH, \qP_{\miPone} , \qD \}. \notag
\end{align}
The general solutions for the $\Omega$-forms are
\begin{align} \label{eq:SchrodOs}
    \Omega_{\mione \mitwo} &= 
    \left( 
    \begin{array}{ccc}
        \Omega_{\qH \qH} & 0 & 0 \\
        0 & 0 & 0 \\
        0 & 0 & 0
    \end{array}
    \right), &
    \Omega^{\mione \mitwo} &=
    \left(
    \begin{array}{ccc}
        0 & 0 & 0 \\
        0 & \Omega^{\qP \qP} I & 0 \\
        0 & 0 & \Omega^{\qD \qD}
    \end{array}
    \right).
\end{align}
Consider the coset representative
\begin{equation} \label{eq:Schrodcorep}
    \Gel = e^{\qH t} e^{\qP_{\miPone} x^{\miPone}} e^{- \qD \ln r}.
\end{equation}
Note that here we have chosen $-\ln r$ to be the dual coordinate to the generator $\qD$; changing to a different function here just amounts to choosing a different coordinate.

The Maurer-Cartan form is
\begin{equation} \label{eq:SchrodMC}
    \Gel^{-1} d \Gel = \qH \frac{dt}{r^2} + \qP_{\miPone} \frac{dx^{\miPone}}{r} - D \frac{dr}{r}.
\end{equation}
This gives the Newton-Cartan data
\begin{align} \label{eq:Schrodmet}
    \tau &= \frac{dt}{r^2}, \notag \\
    h^{\mu\nu} &= r^2 \bigl( \delta_{\miPone}^{\mu} \delta_{\miPone}^{\nu} + \delta_{r}^{\mu} \delta_{r}^{\nu} \bigr), \\
    m &= 0. \notag
\end{align}
Here we set $\Omega^{PP}=\Omega^{DD}$ which is equivalent to setting the radius of the spacetime equal to one.
This is an example of a twistless torsional Newton-Cartan (TTNC) geometry since $\tau \wedge d \tau = 0$, but $d\tau \neq 0$.

The combinations
\begin{equation}
\begin{aligned}
    \Omega_{\mione \mitwo} \viel^{\mione} \viel^{\mitwo} &= \frac{dt^2}{r^4}, \\
    h_{\mu\nu} dx^{\mu} dx^{\nu} &= \frac{dx_{\miPone} dx^{\miPone} + dr^2}{r^2},
\end{aligned}
\end{equation}
describe a Newton-Cartan spacetime with $z=2$ Lifshitz scaling. The form of the NC metric data resembles that of a Lifshitz spacetime. However from the coset construction we should expect that this geometry has more than Lifshitz symmetries. We show further below that this is indeed the case.

In addition to a local Galilean boost, a local $SO(2)$ rotation and $U(1)$ transformation, we have a local special conformal transformation $e^{\qK \sct}$ parametrized by $\sct$. The Newton--Cartan data transforms as
\begin{equation}
\begin{aligned}
    \tau &\rightarrow \tau' = \tau, \\
    \viel^{\miPone} &\rightarrow \viel^{\miPone}{}' = \viel^{\miPone} + \lgb^{\miPone} \tau + \epsilon^{\miPone}{}_{\miPtwo} \lambda \viel^{\miPtwo}, \\
    \viel^r &\rightarrow \viel^r{}' = \viel^r + \sct \tau, \\
    m &\rightarrow m' = m + \lgb_{\miPone} \viel^{\miPone} + \frac{1}{2} \lgb_{\miPone} \lgb^{\miPone} \tau + d \lut.
\end{aligned}
\end{equation}
These are the local coset symmetries obtained by right multiplication of a coset representative by an element of the subgroup $H$. The right multiplication by $e^{K\omega}$ agrees with the action of the local special conformal transformation given in \cite{Bergshoeff:2014uea}. For $\omega=0$ they form a subset of the local TNC symmetries discussed in equation \eqref{eq:TNCsym}, because there all the vielbeins are treated on an equal footing whereas here the vielbein associated with dilatations breaks the local tangent space isotropy from $SO(3)$ down to $SO(2)$.


\subsection{Schr\"odinger via Null Reduction}
\label{sec:SchrodNR}

We cannot derive Schr\"odinger as the contraction of a relativistic algebra, but we can try to recover the previous results via null reduction. However, this will not be as simple as it was for Bargmann in Sec. \ref{sec:flatNCnullreduction} or Newton-Hooke in Sec. \ref{sec:NHNR}. In those cases, all we had to do was move $\qN$ from $\Lieh$ to $\cosetm$. Doing so for Schr\"odinger leads to degenerate $\Omega$-forms, which does not lend itself to null reduction. However, the task of determining the appropriate coset to take with $\qN \in \cosetm$ to produce a \ndeg metric was already completed in~\cite{SchaferNameki:2009xr}. The appropriate split of generators is 
\begin{align} \label{eq:SchrodNRquotient}
    \Lieg &= \{ \qN , \qH, \qP_{\miPone} , \qD , \qJ, \qG_{\miPone}, \qK  \}, \notag \\
    \Lieh &= \{ \qJ, \qG_{\miPone}, 2 \qK + \qN \}, \\
    \qm &= \{ \qN , \qH, \qP_{\miPone} , \qD \}. \notag
\end{align}
We note that this is a non-reductive coset giving rise to a homogeneous pseudo-Riemannian spacetime.
Choosing the coset representative
\begin{equation}
   g =e^{\qH t}e^{\qP_{\miPone}x^\mione}e^{-D \ln r}e^{-Nu},
\end{equation}
leads to the metric
\begin{equation}
    ds^2 = 2 \frac{dt}{r^2} \left( du - \frac{\alpha}{2} \frac{dt}{r^2} \right) + \frac{dx_{\miPone} dx^{\miPone} + dr^2}{r^2},
\end{equation}
where $\alpha$ is some constant. %
We identify the Newton-Cartan data of the null reduction of this spacetime:
\begin{align} \label{eq:SchrodNCNR}
    \tau &= \frac{dt}{r^2}, \notag \\
    h^{\mu\nu} &= r^2 \bigl( \delta_{\miPone}^{\mu} \delta_{\miPone}^{\nu} + \delta_{r}^{\mu} \delta_{r}^{\nu} \bigr), \\
    m &= \frac{\alpha}{2} \frac{dt}{r^2}. \notag
\end{align}
This is almost the same as (\ref{eq:Schrodmet}), but with a nonzero $m$. This time, however, we cannot transform $m$ away using a local Galilean (Milne) boost and a $U(1)$ transformation. How do we reconcile this with the previous Schr\"odinger coset? 

It turns out that there is actually a one-parameter family of Schr\"odinger cosets we can take that all produce Newton-Cartan spacetimes with Schr\"odinger symmetry. The one parameter is precisely $\alpha$ and the coset in Section \ref{sec:Schrodcoset} is the $\alpha = 0$ case. We turn now to this one-parameter family of Schr\"odinger cosets.


\subsection{Newton-Cartan Geometries with Schr\"odinger Symmetry}
\label{sec:NCSchrod}

Instead of (\ref{eq:Schrodquotient}), consider splitting the generators according to
\begin{align} \label{eq:Schrodquotientprime}
    \Lieg &= \{ \overline{\qH}, \qP_{\miPone} , \qD , \qJ, \qG_{\miPone}, \qN, \qK \}, \notag \\
    \Lieh &= \{ \qJ, \qG_{\miPone}, \qN, \qK \}, \\
    \qm &= \{ \overline{\qH}, \qP_{\miPone} , \qD \}, \notag
\end{align}
where
\begin{equation}
    \overline{\qH} = \qH - \frac{\alpha}{2} N.
\end{equation}
Then, (\ref{eq:Schrodquotient}) is just the limit $\alpha = 0$. Now, the commutator between $\overline{\qH}$ and $\qD$ reads
\begin{equation}
    [ \overline{H} , \qD ] = 2 \overline{\qH} + \alpha \qN.
\end{equation}
Thus, we can now produce $\qN$ (and therefore $m$) from the commutator of $\qH$ and $\qD$.

Consider the same coset representative as (\ref{eq:Schrodcorep}), but with $\qH$ replaced with $\overline{\qH}$. Then, the Maurer-Cartan form reads
\begin{equation}
    g^{-1} dg = \overline{H} \frac{dt}{r^2} + \qP_{\miPone} \frac{dx^{\miPone}}{r} - \qD \frac{dr}{r} + \frac{\alpha}{2} N dt \left( \frac{1}{r^2} -1 \right),
\end{equation}
which gives the following Newton-Cartan data
\begin{align}
    \tau &= \frac{dt}{r^2}, \notag \\ \label{eq:SchrGenCoset}
    h^{\mu\nu} &= r^2 \bigl( \delta_{\miPone}^{\mu} \delta_{\miPone}^{\nu} + \delta_{r}^{\mu} \delta_{r}^{\nu} \bigr), \\
    m &= \frac{\alpha}{2} \frac{dt}{r^2} - \frac{\alpha}{2} dt. \notag
\end{align}
The only difference between this and the result of the null reduction (\ref{eq:SchrodNCNR}) is the extra $- \frac{\alpha}{2} dt$ in $m$. That is an exact form and we can always get rid of it by a local $U(1)$ transformation $\sigma = \alpha t / 2$.

We can see that these different Newton-Cartan spacetimes all have the same Schr\"odinger symmetries, namely the same Killing vectors, by the following argument. A Killing vector $\Kill$ (not to be confused with the generator of special conformal transformations) satisfies the equations
\begin{align} \label{eq:Killingeqs}
    \Lie_{\Kill} \tau_{\mu} &= 0, \notag \\
    \Lie_{\Kill} m_{\mu} &= - \lambda_{\mu} - \partial_{\mu} \sigma, \\
    \Lie_{\Kill} h_{\mu\nu} &= - \lambda_{\mu} \tau_{\nu} - \lambda_{\nu} \tau_{\mu}, \notag
\end{align}
where $\sigma$ is some scalar function and $\lambda_{\mu}=\lambda_a \viel^a_\mu$. These equations follow from demanding that the NC data is invariant under a diffeomorphism generated by $\Kill$ up to a local gauge transformation of the form \eqref{eq:TNCsym}. The only dependence on $\alpha$ comes in through the second Killing equation since $m = \frac{\alpha}{2} \frac{dt}{r^2}$ appears explicitly. However, since $m_{\mu} = \frac{\alpha}{2} \tau_{\mu}$, the first equation implies
\begin{equation} \label{eq:Killingm}
    \Lie_{\Kill} m_{\mu} = 0.
\end{equation}
Therefore, the second Killing equation just leads to the identification $\lambda_{\mu} = - \partial_{\mu} \sigma$. Indeed, regardless of $\alpha$, there are eight Killing vectors, which we label below by the symmetry generator which they represent,
\begin{align} \label{eq:SchrodKs}
    H &= \partial_t, &%
    J &= - \epsilon_{i}^{\phantom{i} j} x^i \partial_j, \notag \\
    P_i &= \partial_i, &%
    D &= 2t \partial_t + r \partial_r + x^i \partial_i, \\
    G_i &= t \partial_i, &%
    K &= t ( t \partial_t + r \partial_r + x^i \partial_i ). \notag
\end{align}
There are eight Killing vectors here since $i = 1,2$. Note that $N$ is not realized as a Killing vector, which is consistent with the fact that these spacetimes can be retrieved via null reduction. It is $\partial_{u}$ that would have played the role of the Killing vector associated with $N$, but $u$ is not actually a coordinate after null reduction. As a result, there is no Killing vector associated with $N$. 

Finally, we note that, with the presence of the dilatation operator and its dual radial coordinate in this Schr\"odinger algebra case, we have now constructed Newton-Cartan spacetimes that can be meaningfully considered in the context of holography. 
Specifically, just as the Riemmannian metric (\ref{eq:SchrodNRquotient}) was proposed e.g. in \cite{Son:2008ye} as a locally relativistic holographic dual for a field theory with Schr\"odinger symmetries, here we propose the Newton-Cartan data in (\ref{eq:SchrGenCoset}) be interpreted as an intrinsically \nrel dual for field theories with Schr\"odinger symmetries. We postpone further discussion of this proposal to the discussion section. 


\section{Cosets with \texorpdfstring{$SO(3)$}{SO(3)} isometries and non-Lorentzian geometries}\label{sec:Sphere}

The previous three examples used kinematical algebras, or the closely related Schr\"odinger algebra, which in some sense are immediately well-suited to the coset procedure. In this section, we will show an example which is much less obvious and which uses a non-kinematical algebra.

We wish to generate a Newton-Cartan coset space with the topology $\mathbb{R}\times S^2$, that is a space with some $SO(3)$ isometry. The Hopf fibration of the 3-sphere into what is locally a product of a circle and a 2-sphere is ideal for this purpose: one can null reduce along some combination of the coordinate on the Hopf circle and a separate coordinate along a real line. The resulting Newton-Cartan spacetime should be spatially $S^2$-like. Additionally, we expect to find spaces with generic torsion, i.e. torsional Newton-Cartan (TNC) geometries.

In this case, we have a clearer idea of what we want the null reduction to look like compared to what the coset procedure should look like. Therefore, we will reverse the logic of the presentation compared to the previous examples and consider the appropriate null reduction first before determining the corresponding coset. The appropriate coset procedure will become transparent once we have determined the Killing vectors of the geometry that we wish to construct.

\subsection{The unit 3-sphere}

Consider the unit 3-sphere embedded in $\mathbb{R}^4$
\begin{equation}
(x^1)^2+(x^2)^2+(x^3)^2+(x^4)^2=1\,.
\end{equation}
We can parametrize this in terms of hyperspherical coordinates as
\begin{equation}
\begin{aligned}
x^1 & = \cos\chi\,,\\
x^2 & = \sin\chi\cos\theta\,,\\
x^3 & = \sin\chi\sin\theta\cos\varphi\,,\\
x^4 & = \sin\chi\sin\theta\sin\varphi\,,
\end{aligned}
\end{equation}
where $0\le\chi\le\pi$, $0\le\theta\le\pi$ and $0\le\varphi<2\pi$. The metric induced on the 3-sphere is 
\begin{equation}
ds^2_3=\gamma_{ij}dx^idx^j=d\chi^2+\sin^2\chi\left(d\theta^2+\sin^2\theta d\varphi^2\right)\,.
\end{equation}
An alternative coordinate system in terms of Hopf coordinates can be obtained by the following parametrization
\begin{equation}
\begin{aligned}
x^1 & = \cos\phi_+\sin\psi\,,\\
x^2 & = \sin\phi_+\sin\psi\,,\\
x^3 & = \cos\phi_-\cos\psi\,,\\
x^4 & = \sin\phi_-\cos\psi\,,
\end{aligned}
\end{equation}
where $0\le\psi\le\pi/2$, $0\le\phi_+<2\pi$ and $0\le\phi_-<2\pi$. The metric on the unit 3-sphere in these coordinates is given by
\begin{equation}
ds^2=d\psi^2+\sin^2\psi d\phi_+^2+\cos^2\psi d\phi_-^2\,.
\end{equation}
This coordinate system makes manifest the presence of two commuting $S^1$ isometries whose Killing vectors are $\partial_{\phi_+}$ and $\partial_{\phi_-}$.

If we define the coordinates $\phi_R$, $\phi_L$ and $\eta$ via
\begin{equation}
\phi_R=\phi_++\phi_-\,,\qquad \phi_L=\phi_+-\phi_-\,,\qquad \eta=2\psi\,,
\end{equation}
then the metric becomes
\begin{equation}\label{eq:S3metric}
ds^2=\frac{1}{4}\left(d\eta^2+d\phi_R^2+d\phi_L^2-2\cos\eta \, d\phi_R \, d\phi_L\right)\,.
\end{equation}
This can be written as a Hopf fibration of $S^1$ over $S^2$
\begin{equation}
ds^2=\frac{1}{4}\left(d\eta^2+\sin^2\eta d\phi_R^2+\left(d\phi_L -\cos\eta \, d\phi_R\right)^2\right)\,.
\end{equation}

The notation $\phi_L$ and $\phi_R$ is adapted to the isometry between the Lie algebra $\mathfrak{so}(4)$ and two copies of $\mathfrak{su}(2)$, a left and a right $\mathfrak{su}(2)$, whose $U(1)$ Cartan generators correspond to $\partial_R=\partial_{\phi_R}$ and $\partial_L=\partial_{\phi_L}$.

The Killing vectors of the 3-sphere in the coordinates \eqref{eq:S3metric} are given by
\begin{equation}
\begin{aligned}
J_{1}^L & = \sin\phi_L\partial_\eta+\frac{\cos\phi_L}{\sin\eta}\partial_R+\cos\phi_L\cot\eta \, \partial_L\,,\\
J_{2}^L & = \cos\phi_L\partial_\eta-\frac{\sin\phi_L}{\sin\eta}\partial_R-\sin\phi_L\cot\eta \, \partial_L\,,\\
J_{3}^L & = \partial_L\,,\label{eq:JLz}\\
J_{1}^R & = \sin\phi_R\partial_\eta+\cos\phi_R\cot\eta \, \partial_R+\frac{\cos\phi_R}{\sin\eta}\partial_L\,,\\
J_{2}^R & = \cos\phi_R\partial_\eta-\sin\phi_R\cot\eta \, \partial_R-\frac{\sin\phi_R}{\sin\eta}\partial_L\,,\\
J_{3}^R & = \partial_R\,.
\end{aligned}
\end{equation}
The left generators $J_{a}^L$ and the right generators $J_{a}^R$ where $a=1,2,3$ are related by interchanging $\phi_L$ and $\phi_R$. They satisfy the Lie algebra $\mathfrak{su}(2)_L\oplus\mathfrak{su}(2)_R$ where 
\begin{equation}
\left[J_{a}^L\,,J_{b}^L\right]=\epsilon_{abc}J_{c}^L\,,
\end{equation}
with $\epsilon_{123}=1$ for $\mathfrak{su}(2)_L$ and 
\begin{equation}
\left[J_{a}^R\,,J_{b}^R\right]=\epsilon_{abc}J_{c}^R\,,
\end{equation}
for $\mathfrak{su}(2)_R$. Each $\mathfrak{su}(2)_L$ generator commutes with each $\mathfrak{su}(2)_R$ generator. This is of course due to the fact that the Lie algebra $\mathfrak{so}(4)$ is isomorphic to $\mathfrak{su}(2)_L\oplus\mathfrak{su}(2)_R$.

\subsection{TNC geometries with \texorpdfstring{$\mathbb{R}\times SO(3)$}{RxSO(3)} isometries}

Consider the Lorentzian metric describing a line times a unit 3-sphere:
\begin{equation}\label{eq:RxS3}
ds^2=-dt^2+\frac{1}{4}\left[d\eta^2+\sin^2\eta \, d\phi_R^2+(d\phi_L-\cos\eta \, d\phi_R)^2\right]\,.
\end{equation}
This metric can be written in the form of a null reduction by defining the coordinates $u$ and $v$ via
\begin{equation}\label{eq:coordinatetrafo}
u=\frac{1}{4}\phi_L-\frac{1}{2}t\,,\qquad v=\frac{1}{2}\phi_L+t\,.
\end{equation}
The metric \eqref{eq:RxS3} becomes
\begin{equation}
ds^2=2\tau(du-m)+h_{\mu\nu}dx^\mu dx^\nu\,,
\end{equation}
where
\begin{equation}\label{eq:TNCwithSO(3)}
\tau=dv-\frac{1}{2}\cos\eta \, d\phi_R\,,\qquad m=\frac{1}{4}\cos\eta \, d\phi_R\,,\qquad h_{\mu\nu}dx^\mu dx^\nu=\frac{1}{4}\left(d\eta^2+\sin^2\eta \, d\phi_R^2\right)\,.
\end{equation}

The original metric \eqref{eq:RxS3} has 7 isometries in 4 spacetime dimensions. These are $\partial_t$ and the 6 Killing vectors of the 3-sphere. What are the Killing vectors of the null reduced geometry? Clearly any Killing vector of the null reduced TNC geometry must be a Killing vector of the uplifted Lorentzian spacetime. A TNC Killing vector $\Kill^\mu$ is defined so that the action of the Lie derivative along $\Kill$ acting on $\tau_\mu$, $m_\mu$ and $h_{\mu\nu}$ is zero up to a local TNC gauge symmetry (redundancy). Since the latter leave the null reduction ansatz invariant any TNC Killing vector is also a Killing vector of the higher-dimensional Lorentzian metric. 

The coordinate transformation \eqref{eq:coordinatetrafo} from $(t,\phi_L)$ to $(u,v)$ implies
\begin{equation}
\partial_L=\frac{1}{2}\partial_v+\frac{1}{4}\partial_u\,,\qquad\partial_t=\partial_v-\frac{1}{2}\partial_u\,.
\end{equation}
A quick way to get the Killing vectors of the null reduced geometry is to apply the coordinate transformation \eqref{eq:coordinatetrafo} to the 7 Killing vectors of the Lorentzian metric and to throw away all Killing vectors that depend explicitly on $u$, i.e. do not commute with $\partial_u$. This leaves us with 4 Killing vectors, namely $\partial_t$ and the $\mathfrak{su}(2)_R$ sector. After applying the coordinate transformation \eqref{eq:coordinatetrafo} these will contain terms proportional to $\partial_u$. The vector $\partial_u$ does not exist in the reduced geometry because $u$ is not a coordinate in the lower-dimensional geometry. Fortunately all terms proportional to $\partial_u$ can be dropped because they generate local TNC symmetries corresponding to Bargmann gauge transformations. This leaves us with the following 4 Killing vectors
\begin{equation}
\begin{aligned}
H & = \partial_v\,,\\
J_{1} & = \sin\phi_R \, \partial_\eta+\cos\phi_R\cot\eta \, \partial_R+\frac{1}{2}\frac{\cos\phi_R}{\sin\eta}\partial_v\,,\\
J_{2} & = \cos\phi_R \, \partial_\eta-\sin\phi_R\cot\eta \, \partial_R-\frac{1}{2}\frac{\sin\phi_R}{\sin\eta}\partial_v\,,\\
J_{3} & = \partial_R\,.
\end{aligned}
\end{equation}
It can be checked that all four of these Killing vectors $\Kill\in\{H,J_1,J_2,J_3\}$ obey
\begin{equation}
\mathcal{L}_\Kill\tau_\mu=0\,,\qquad\mathcal{L}_\Kill h_{\mu\nu}=0\,,\qquad \mathcal{L}_\Kill m_\mu=\partial_\mu\sigma\,,
\end{equation}
where $\sigma$ is given by
\begin{equation}
\sigma=\frac{1}{4}\frac{\cos\phi_R}{\sin\eta}
\end{equation}
for the Killing vector $K=J_1$ and by
\begin{equation}
\sigma=-\frac{1}{4}\frac{\sin\phi_R}{\sin\eta}
\end{equation}
for the Killing vector $K=J_2$ and where $\sigma=0$ for the other two Killing vectors $H$ and $J_3$. Note that we did not have to use the TNC redundancy with respect to local Galilean boosts. The Lie algebra spanned by $\{J_1,J_2,J_3\}$ is that of $\mathfrak{so}(3)$ while $H$ commutes with all the $J_a$ generators. 

Are there other TNC geometries with the same algebra of Killing vectors? Looking at \eqref{eq:TNCwithSO(3)} we can see that the following Killing vectors 
\begin{equation}
\begin{aligned}
H & = \partial_v\,,\\
J_{1} & = \sin\phi_R \, \partial_\eta+\cos\phi_R\cot\eta \, \partial_R\,,\\
J_{2} & = \cos\phi_R \,  \partial_\eta-\sin\phi_R\cot\eta \, \partial_R\,,\\
J_{3} & = \partial_R\,.
\end{aligned}
\end{equation}
where we removed the $\partial_v$ terms in the expressions for $J_1$ and $J_2$ are isometries of the following TNC geometry
\begin{equation}\label{eq:TNCwithSO(3)-2}
\tau=dv\,,\qquad m=0\,,\qquad h_{\mu\nu}dx^\mu dx^\nu=d\eta^2+\sin^2\eta \, d\phi_R^2\,.
\end{equation}
This is the TNC geometry that one would obtain from the null reduction of $\mathbb{R}^{1,1}\times S^2$ with metric
\begin{equation}
ds^2=2dvdu+d\eta^2+\sin^2\eta \, d\phi_R^2\,.
\end{equation}

\subsection{Coset descriptions}

In order to find coset descriptions of the above TNC geometries let us define the generators $H$, $J_1$, $J_2$, $J_3$ and $N$ where $H$ and $N$ commute with every other generator and the $J_a$ form an $\mathfrak{so}(3)$ algebra. In order to study spheres of different radii we define
\begin{equation}
    P_1 = \frac{1}{R} J_1\,, \qquad
    P_2 = \frac{1}{R} J_2\,, \qquad
    J = J_3\,.
\end{equation}
where $J_1$ and $J_2$ have been rescaled by $R$, the radius, so that we get the nonzero commutators 
\begin{equation}
[P_1,P_2]=\frac{1}{R^2}J\,,\qquad [J,P_1]=P_2\,,\qquad [J,P_2]=-P_1\,.
\end{equation}
We can think of $N$ as the generator $\partial_u$ in the null reduction. This is a symmetry that is not realized as a Killing vector in the lower-dimensional setting, but that will play an important role in defining the TNC gauge field $m$. We split the algebra as follows:
\begin{equation}
    \Lieh = \{ X= J + aN + bH , N \}\,, \qquad
    \cosetm = \{ H, P_1, P_2 \}\,.
\end{equation}
The most general invariant metrics on the coset are
\begin{equation}
\Omega_{ab}=\text{diag}(\Omega_{HH},\Omega_{PP},\Omega_{PP})\,,\qquad \Omega^{ab}=\text{diag}(\Omega^{HH},\Omega^{PP},\Omega^{PP})\,,
\end{equation}
where $a=H,P_1,P_2$ and where $\Omega_{P_1P_1}=\Omega_{P_2P_2}=\Omega_{PP}$ and $\Omega^{P_1P_1}=\Omega^{P_2P_2}=\Omega^{PP}$. This result means that this coset is compatible with different signature choices for $\Omega_{ab}$. We can take the signature $(1,0,0)$ like in TNC geometry by setting $\Omega_{PP}=0$ or we can take the signature $(0,1,1)$ like in Carrollian geometry by setting $\Omega_{HH}=0$. But we can also take Riemannian and pseudo-Riemannian signatures by setting $\Omega_{HH}=\pm\Omega_{PP}$.

Let us take the following coset representative
\begin{equation}
g=e^{Hv}e^{P_1\alpha_1}e^{P_2\alpha_2}\,.
\end{equation}
The Maurer--Cartan 1-form is expanded as
\begin{equation}
g^{-1}dg=H\tau+P_1 \viel^1+P_2 \viel^2+Nm+X\omega\,.
\end{equation}
Using 
\begin{equation}
e^{-P_2\alpha_2}P_1e^{P_2\alpha_2}=P_1\cos\frac{\alpha_2}{R}+\frac{1}{R}J \sin\frac{\alpha_2}{R}\,,
\end{equation}
we then find
\begin{equation}
\begin{aligned}
    \tau &= dv-\frac{b}{R}\sin\frac{\alpha_2}{R}d\alpha_1\,, &%
    \viel^1 &=\cos\frac{\alpha_2}{R}d\alpha_1\,, &%
    \viel^2 &= d\alpha_2\,, \\
    m &=-\frac{a}{R}\sin\frac{\alpha_2}{R}d\alpha_1\,, &%
    \omega &=\frac{1}{R}\sin\frac{\alpha_2}{R}d\alpha_1\,.
\end{aligned}
\end{equation}
Next we make a simple coordinate transformation
\begin{equation}
\alpha_1=R\phi_R\,,\qquad\alpha_2=R \Bigl( \eta-\frac{\pi}{2} \Bigr)\,,
\end{equation}
so that we obtain
\begin{equation}
\begin{aligned}
    \tau &= dv+b\cos\eta \, d\phi_R\,, &%
    \viel^1 &= R\sin\eta \, d\phi_R\,, &%
    \viel^2 &= Rd\eta\,,\\
    m &=a\cos\eta \, d\phi_R\,, &%
    \omega &= -\cos\eta \, d\phi_R\,.
\end{aligned}
\end{equation}
We thus find that upon setting $a=\frac{1}{4}$, $b=-\frac{1}{2}$ and $R=\frac{1}{2}$ we reproduce the TNC geometry \eqref{eq:TNCwithSO(3)}, whereas if we set $a=b=0$ and $R=1$ we obtain the TNC geometry \eqref{eq:TNCwithSO(3)-2}. This of course requires that we choose the metrics $\Omega_{ab}=\text{diag}(1,0,0)$ and $\Omega^{ab}=\text{diag}(0,1,1)$.

The results of this section thus show that the coset procedure also allows to obtain non-trivial TNC geometries. In particular our example features a coset of
a non-kinematical algebra, giving rise to a geometry that includes spherical topology in the spatial directions.
Moreover, this example illustrates that for given $\GroupG$ and $\GroupH$ there can be situations where
the invariant bilinear form on the coset is not necessarily degenerate, but by choosing it to be so one
can arrive at interesting NC geometries. We finally note that the TNC spacetimes with $SO(3)$ isometries that we have constructed herein are spacetimes that have appeared in the context of
nonrelativistic strings and limits of the AdS/CFT correspondence~\cite{Harmark:2017rpg}.


\section{Discussion and Outlook}\label{sec:Discussion}

We conclude with a brief discussion on some generalizations and applications. 

First of all, there are many interesting directions in which to extend the coset construction of Newton-Cartan spacetimes described in this paper. Some of these directions ought to be straightforward; for example, one could begin to catalog all possible Newton-Cartan coset spacetimes. One could also consider supersymmetric extensions of these \nrel algebras, such as the supersymmetric Newton-Hooke algebras considered in \cite{Bergshoeff:2014gja}, which should in principle generate Newton-Cartan superspace. More generally, applying
the construction to other kinematical algebras could also be worthwhile to consider.\footnote{See Ref.~\cite{Figueroa-OFarrill:2017sfs,Figueroa-OFarrill:2017ycu,Figueroa-OFarrill:2017tcy} for a recent classification of kinematical algebras in general dimensions; for the seminal work in $3+1$ dimensions, see \cite{Bacry:1968zf}.} Coset prescriptions could also be of use in exploring higher spin geometries.
We also note that quotienting maximally symmetric NC cosets
in three dimensions by discrete subgroups, one might arrive at  non-Lorentzian analogues of the BTZ black holes.

Another direction to pursue is to see whether the coset spaces we obtained
(along with possible others) are natural solutions of Ho\v{r}ava-Lifshitz gravity and related \nrel Chern-Simons theories \cite{Hartong:2015zia,Hartong:2016yrf}. 
As a concrete example, in Ref. \cite{Hartong:2017bwq} it will be shown that the vacuum of a novel \nrel CS theory,
that can be obtained from a limit of $SO(2,2)+U(1)+U(1)$ CS theory by zooming in on a BPS bound, is described by the generalized coset prescription presented in this paper. In that case the relevant bulk geometry is a non-Lorentzian geometry called (pseudo)-Newton Cartan geometry.
Furthermore, since NC geometry also appears as the target space in \nrel string theory \cite{Harmark:2017rpg} there might be useful applications of the coset structure in that context.%
\footnote{Coset constructions of \nrel algebras have been considered in \nrel worldline and sigma-model actions, see e.g. Refs.~\cite{Gomis:2005pg,Brugues:2006yd,Bergshoeff:2014gja}, 
albeit in a seemingly different way.} 

As briefly mentioned at the end of Section \ref{sec:Schrodinger}, we believe the Newton-Cartan data we find in (\ref{eq:SchrGenCoset}) should be taken seriously as a proposal for an intrinsically \nrel dual of Schr\"odinger field theories. Since these spaces can be obtained from null reductions of spaces like (\ref{eq:SchrodNRquotient}), the distinction may seem somewhat academic; however the local structure of Newton-Cartan spaces such as (\ref{eq:SchrGenCoset}) materially differs from that of their null lifts.  One first step in studying these proposed duals would be to compare with existing alternate duals, such as the lower spin gravity proposed in \cite{Hofman:2014loa}.\footnote{We thank Wei Song for pointing out this possibility.}
Additionally, if these structures do turn out to be fruitful for studying Schr\"odinger holography, we may then ask if other symmetry algebras that include a scaling generator, such as Lifshitz, can produce similarly interesting holographic proposals.

 Turning to another potential application, we believe the coset structures described herein may be useful for studying entanglement entropies, along the lines of \cite{Castro:2015csg}. We note recent work indicates that the entanglement structure of the Galilean vacuum is trivial~\cite{Hason:2017flq}. The argument makes crucial use of the existence of the $U(1)$ (particle number) symmetry generator $N$. It is worth emphasizing here that this should not actually discourage one from considering the question of how to generalize the Ryu-Takayanagi prescription to cases with \nrel bulk geometries. In fact, particle number conservation is generically broken when one considers interesting cases. For example, the dual of a black brane moving in a Lifshitz bulk spacetime, studied in \cite{Hartong:2016nyx}, is a Lifshitz perfect fluid on a Newton-Cartan spacetime arising from a Schr\"odinger perfect fluid with broken particle number symmetry.  Additionally, not every \nrel theory necessarily has Galilean symmetry, for example Lifshitz scaling symmetry need not include Galilean generators.%
\footnote{See Refs.~\cite{MohammadiMozaffar:2017nri,He:2017wla,Gentle:2017ywk} for recent work on entanglement entropy in Lifshitz field theories and  holographic proposals.}

In addition to these physical considerations, there are also several questions of mathematical interest.  Most mathematical works assume a \ndeg $\Omega_{ab}$ is available.  Discovering what properties of coset structures still hold for cases when $\Omega_{ab}$ and $\Omega^{ab}$ are degenerate is a mathematically interesting question, as well as having obvious physical import.



\section*{Acknowledgements}

We would like to thank Eric Bergshoeff, Jaume Gomis, Diego Hofman, Yang Lei, Gerben Oling, Sakura Schafer-Nameki and Wei Song for useful discussions. 
The authors gratefully acknowledge support from the Simons Center for Geometry and Physics, Stony Brook University at which some of the research for this paper was performed during the 2017 workshop ``Applied Newton-Cartan Geometry''. J.H. acknowledges hospitality of Niels Bohr Institute. 
 The work of K.T.G. is supported by the ERC Advanced Grant 291092 ``Exploring the Quantum Universe." 
 The  work  of  J.H.  was  in  part
supported by a STSM Grant from COST Action MP1405 QSPACE.
  The work of C.K. was partially supported by the European Union's Horizon 2020 research and innovation programme under the Marie Sk\l{}odowska-Curie grant agreement No 656900. 
  The work of N.O. is supported in part by 
the project ``Towards a deeper understanding of  black holes with \nrel holography'' of the Independent Research Fund Denmark
(grant number DFF-6108-00340)
  
\begin{appendix}

\section{Aristotelian Algebra}
\label{sec:ARGapp}

The Aristotelian algebra consists of the Hamiltonian $\qH$, spatial translations $\qP_{\miPone}$ and spatial rotations $\qJ_{\miPone \miPtwo}$,
\begin{equation} \label{eq:argens}
    \qg = \{ \qH, \qP_{\miPone}, \qJ_{\miPone \miPtwo} \}.
\end{equation}
The algebra is given by
\begin{equation} \label{eq:aralg}
\begin{aligned}[]
    [\qJ_{\miPone \miPtwo}, \qJ_{\miPthree \miPfour}] &= \delta_{\miPone \miPthree} \qJ_{\miPtwo \miPfour} - \delta_{\miPone \miPfour} \qJ_{\miPtwo \miPthree} - \delta_{\miPtwo \miPthree} \qJ_{\miPone \miPfour} + \delta_{\miPtwo \miPfour} \qJ_{\miPone \miPthree}, \\
    [\qJ_{\miPone \miPtwo}, \qP_{\miPthree} ] &= \delta_{\miPone \miPthree} \qP_{\miPtwo} - \delta_{\miPtwo \miPthree} \qP_{\miPone}.
\end{aligned}
\end{equation}
We quotient out by the subalgebra of spatial rotations,
\begin{equation} \label{eq:arhandm}
\begin{aligned}[]
    \qh &= \{ \qJ_{\miPone \miPtwo} \}, \\
    \qm &= \{ \qH, \qP_{\miPone} \}.
\end{aligned}
\end{equation}
The general solutions for the $\Omega$ forms in \eqref{eq:Omegalower} and \eqref{eq:Omegaupper} are
\begin{align} \label{eq:arOs}
    \Omega_{\mione \mitwo} &= 
    \left( 
    \begin{array}{cc}
        \Omega_{\qH \qH} & 0 \\
        0 & \Omega_{\qP \qP} I 
    \end{array}
    \right), &
    \Omega^{\mione \mitwo} &=
    \left(
    \begin{array}{cc}
        \Omega^{\qH \qH} & 0 \\
        0 & \Omega^{\qP \qP} I
    \end{array}
    \right),
\end{align}
where $I$ is the identity matrix in the spatial directions. If $\Omega^{\qH \qH} = \Omega_{\qH \qH}^{-1}$ and $\Omega^{\qP \qP} = \Omega_{\qP \qP}^{-1}$, then these two matrices would be mutual inverses.

Since $[ \mathfrak{m} , \mathfrak{m} ] = 0$, the Maurer-Cartan form is simply $\qH \, dt + \qP_{\miPone} \, dx^{\miPone}$, where $t$ and $x^{\miPone}$ are the coordinates associated with $\qH$ and $\qP_{\miPone}$, respectively. The metric is \ndegafternoun and is simply equal to $\Omega$ (after we identify the coordinate with the corresponding generator, e.g. $g_{tt} = \Omega_{\qH \qH}$). One may simply rescale coordinates to set $| \Omega_{HH} | = | \Omega_{PP} | = 1$. Then, the metric is flat with any particular choice of signature.

We can choose $\Omega$ forms with the correct signature to produce Newton-Cartan geometry, namely $\Omega_{\qP \qP} = \Omega^{\qH \qH} = 0$. However, the result would still not quite be Newton-Cartan geometry since there is no object to play the role of the gauge field $\gaugef$ because the algebra lacks the central element $\qN$.


\section{Galilei Algebra}
\label{sec:Galapp}

The Galilei algebra consists of the Hamiltonian $\qH$, spatial translations $\qP_{\miPone}$, spatial rotations $\qJ_{\miPone \miPtwo}$ and Galilean boosts $\qG_{\miPone}$,
\begin{equation} \label{eq:Galgens}
    \qg = \{ \qH, \qP_{\miPone}, \qJ_{\miPone \miPtwo}, \qG_{\miPone} \}.
\end{equation}
The algebra is given by (\ref{eq:aralg}) plus
\begin{equation} \label{eq:Galalg}
\begin{aligned}[]
    [\qJ_{\miPone \miPtwo}, \qG_{\miPthree} ] &= \delta_{\miPone \miPthree} \qG_{\miPtwo} - \delta_{\miPtwo \miPthree} \qG_{\miPone}, \\
    [ \qH , \qG_{\miPone} ] &= \qP_{\miPone}.
\end{aligned}
\end{equation}
Again, since there is no central element, there is nothing to play the role of the gauge field $\gaugef$ and the resulting geometry cannot be Newton-Cartan. 


\subsection{Galilei Coset}
\label{sec:directGalcoset}

We quotient out by the subalgebra of spatial rotations and Galilean boosts,
\begin{equation} \label{eq:Galhandm}
\begin{aligned}[]
    \qh &= \{ \qJ_{\miPone \miPtwo}, \qG_{\miPone} \}, \\
    \qm &= \{ \qH, \qP_{\miPone} \}.
\end{aligned}
\end{equation}
The general solutions for the $\Omega$ forms in \eqref{eq:Omegalower} and \eqref{eq:Omegaupper} are
\begin{align} \label{eq:GalOs}
    \Omega_{\mione \mitwo} &= \Omega_{\qH \qH}
    \left( 
    \begin{array}{cc}
        1 & 0 \\
        0 & 0 
    \end{array}
    \right), &
    \Omega^{\mione \mitwo} &= \Omega^{\qP \qP}
    \left(
    \begin{array}{cc}
        0 & 0 \\
        0 & I
    \end{array}
    \right),
\end{align}
where $I$ is the identity matrix in the spatial directions. Again, we can set $\Omega_{HH}$ and $\Omega^{PP}$ to unit magnitude. 

Neither form is invertible. Therefore, the would-be metric is degenerate. Instead, we have a metric complex,
\begin{equation} \label{eq:Galcomplex}
\begin{aligned}[]
    \tau &= dt, \\
    h^{\mu\nu} &= \delta_{\miPone}^{\mu} \delta_{\miPone}^{\nu}.
\end{aligned}
\end{equation}
%

\subsection{Galilei as Contracted Poincar\'e}
\label{sec:G=cP}

The Galilei algebra can be thought of as an In\"on\"u-Wigner contraction of the Poincar\'e algebra, which is generated by Lorentz transformations $\qrJ_{\mione \mitwo}$, which include boosts, and spacetime translations, $\qrP_{\mione}$, with commutators
\begin{equation} \label{eq:Poincarealg}
\begin{aligned} 
    \phantom{i} [ \qrJ_{\mione \mitwo}, \qrJ_{\mithree \mifour} ] &= \eta_{\mione \mithree} \qrJ_{\mitwo \mifour} - \eta_{\mione \mifour} \qrJ_{\mitwo \mithree} - \eta_{\mitwo \mithree} \qrJ_{\mione \mifour} + \eta_{\mitwo \mifour} \qrJ_{\mione \mithree}, \\
    [\qrJ_{\mione \mitwo} , \qrP_{\mithree}] &= \eta_{\mione \mithree} \qrP_{\mitwo} - \eta_{\mitwo \mithree} \qrP_{\mione}.
\end{aligned}
\end{equation}
where $\eta_{\mu\nu}$ is the flat Minkowski metric with mostly positive signature. We identify
\begin{equation} \label{eq:IWcont}
\begin{aligned}[]
    \qJ_{\miPone \miPtwo} &= \qrJ_{\miPone \miPtwo}, \\
    \qG_{\miPone} &= \frac{1}{c} \qrJ_{0 \miPone}, \\
    \qH &= \qc \qrP_0, \\
    \qP_{\miPone} &= \qrP_{\miPone}.
\end{aligned}
\end{equation}
The algebra is the same as the Galilei algebra, \eqref{eq:aralg} plus \eqref{eq:Galalg}, but with one additional commutator between $\qP$ and $\qG$ and another between $\qG$ and itself: 
\begin{equation} \label{eq:PGcomm}
\begin{aligned}[]
    [\qP_{\miPone}, \qG_{\miPtwo}] &= \frac{1}{c^2} \qH \delta_{\miPone \miPtwo}, \\
    [\qG_{\miPone}, \qG_{\miPtwo}] &= - \frac{1}{c^2} \qJ_{\miPone \miPtwo}.
\end{aligned}
\end{equation}
One recovers the Galilei algebra in the limit $\qc \rightarrow \infty$ and this limit commutes with the coset procedure. The bilinear forms in the case of finite $\qc$ are
\begin{align} \label{eq:Osb4contract}
    \Omega_{\mione \mitwo} &= \Omega_{\qH \qH}
    \left( 
    \begin{array}{cc}
        1 & 0 \\
        0 & - \qc^{-2} I 
    \end{array}
    \right), &
    \Omega^{\mione \mitwo} &= \Omega^{\qP \qP}
    \left(
    \begin{array}{cc}
        - \qc^{-2} & 0 \\
        0 & I
    \end{array}
    \right).
\end{align}
In this way, one can view the degenerate bilinear forms (\ref{eq:GalOs}) as coming from a singular limit of the usual \ndeg relativistic case (\ref{eq:Osb4contract}).

\section{Carroll Algebra}
\label{sec:Carapp}

The Carroll algebra consists of the Hamiltonian $\qH$, spatial translations $\qP_i$, spatial rotations $\qJ_{ij}$ and Carroll boosts $\qC_i$,
\begin{equation} \label{eq:Cargens}
    \qg = \{ \qH, \qP_i , \qJ_{ij} , \qC_i \}.
\end{equation}
The algebra is given by
\begin{align} \label{eq:Caralg}
    [\qJ_{ij}, \qJ_{k \ell}] &= \delta_{ik} \qJ_{j \ell} - \delta_{i \ell} \qJ_{jk} - \delta_{jk} \qJ_{i \ell} + \delta_{j \ell} \qJ_{ik}, \notag \\
    [\qJ_{ij}, \qP_k ] &= \delta_{ik} \qP_j - \delta_{jk} \qP_i, \\
    [\qJ_{ij}, \qC_k ] &= \delta_{ik} \qC_j - \delta_{jk} \qC_i, \notag \\
    [\qP_i , \qC_j] &= \qH \delta_{ij}. \notag
\end{align}
We quotient out by the subalgebra of spatial rotations and Carroll boosts,
\begin{equation} \label{eq:Carhandm}
\begin{aligned}[]
    \qh &= \{ \qJ_{ij}, \qC_i \}, \\
    \qm &= \{ \qH, \qP_i \}.
\end{aligned}
\end{equation}
The general solutions for the $\Omega$-forms are
\begin{align} \label{eq:CarOs}
    \Omega_{\mione \mitwo} &= \Omega_{\qP \qP}
    \left( 
    \begin{array}{cc}
        0 & 0 \\
        0 & I 
    \end{array}
    \right), &
    \Omega^{\mione \mitwo} &= \Omega^{\qH \qH}
    \left(
    \begin{array}{cc}
        1 & 0 \\
        0 & 0
    \end{array}
    \right),
\end{align}
where $I$ is the identity matrix in the spatial directions. 

Scaling $\Omega^{HH}$ and $\Omega_{PP}$ to $\pm 1$ gives the Carroll metric complex
\begin{equation} \label{eq:Carcomp}
\begin{aligned}[]
    v^{\mu} &= \delta_{t}^{\mu}, \\
    h_{\mu\nu} &= \delta_{\mu}^{i} \delta_{\nu}^{i}.
\end{aligned}
\end{equation}
One can also understand this as the $c \rightarrow 0$ limit of the In\"on\"u-Wigner contraction of the Poincar\'e algebra in (\ref{eq:Poincarealg}). In this case, one writes (\ref{eq:Osb4contract}) as
\begin{align} \label{eq:Osb4Carcontract}
    \Omega_{\mione \mitwo} &= \Omega_{\qP \qP}
    \left( 
    \begin{array}{cc}
        - \qc^2 & 0 \\
        0 & I 
    \end{array}
    \right), &
    \Omega^{\mione \mitwo} &= \Omega^{\qH \qH}
    \left(
    \begin{array}{cc}
        1 & 0 \\
        0 & - \qc^2 I
    \end{array}
    \right),
\end{align}
and the $\qc \rightarrow 0$ limit indeed reproduces (\ref{eq:CarOs}).

One can also get Carroll from Bargmann by interchanging the roles of $\qH$ and $\qN$. In this case, the coordinate associated with $\qN$, namely $\qu$, plays the role of time and it is more appropriate to write $\Omega^{\qN \qN}$ in (\ref{eq:CarOs}) instead of $\Omega^{\qH \qH}$ and $v^{\mu} \propto \delta_{u}^{\mu}$ in (\ref{eq:Carcomp}) instead of $\delta_{t}^{\mu}$.

\end{appendix}


\addcontentsline{toc}{section}{References}

\providecommand{\href}[2]{#2}\begingroup\raggedright\endgroup


\begin{thebibliography}{10}

\bibitem{Christensen:2013lma}
M.~H. Christensen, J.~Hartong, N.~A. Obers, and B.~Rollier, ``{Torsional
  Newton-Cartan Geometry and Lifshitz Holography},''
  \href{http://dx.doi.org/10.1103/PhysRevD.89.061901}{{\em Phys.Rev.}
  {\bfseries D89} (2014) 061901},
\href{http://arxiv.org/abs/1311.4794}{{\ttfamily arXiv:1311.4794 [hep-th]}}.

\bibitem{Christensen:2013rfa}
M.~H. Christensen, J.~Hartong, N.~A. Obers, and B.~Rollier, ``{Boundary
  Stress-Energy Tensor and Newton-Cartan Geometry in Lifshitz Holography},''
  \href{http://dx.doi.org/10.1007/JHEP01(2014)057}{{\em JHEP} {\bfseries 1401}
  (2014) 057},
\href{http://arxiv.org/abs/1311.6471}{{\ttfamily arXiv:1311.6471 [hep-th]}}.

\bibitem{Hartong:2014oma}
J.~Hartong, E.~Kiritsis, and N.~A. Obers, ``Lifshitz space--times for
  Schr\"odinger holography,''
  \href{http://dx.doi.org/10.1016/j.physletb.2015.05.010}{{\em Phys. Lett.}
  {\bfseries B746} (2015) 318--324},
\href{http://arxiv.org/abs/1409.1519}{{\ttfamily arXiv:1409.1519 [hep-th]}}.

\bibitem{Hofman:2014loa}
D.~M. Hofman and B.~Rollier, ``{Warped Conformal Field Theory as Lower Spin
  Gravity},'' \href{http://dx.doi.org/10.1016/j.nuclphysb.2015.05.011}{{\em
  Nucl. Phys.} {\bfseries B897} (2015) 1--38},
\href{http://arxiv.org/abs/1411.0672}{{\ttfamily arXiv:1411.0672 [hep-th]}}.

\bibitem{Hartong:2015usd}
J.~Hartong, ``{Holographic Reconstruction of 3D Flat Space-Time},''
  \href{http://dx.doi.org/10.1007/JHEP10(2016)104}{{\em JHEP} {\bfseries 10}
  (2016) 104},
\href{http://arxiv.org/abs/1511.01387}{{\ttfamily arXiv:1511.01387 [hep-th]}}.

\bibitem{Jensen:2017tnb}
K.~Jensen, ``{Locality and anomalies in warped conformal field theory},''
  \href{http://dx.doi.org/10.1007/JHEP12(2017)111}{{\em JHEP} {\bfseries 12}
  (2017) 111},
\href{http://arxiv.org/abs/1710.11626}{{\ttfamily arXiv:1710.11626 [hep-th]}}.

\bibitem{Son:2013rqa}
D.~T. Son, ``{Newton-Cartan Geometry and the Quantum Hall Effect},''
\href{http://arxiv.org/abs/1306.0638}{{\ttfamily arXiv:1306.0638
  [cond-mat.mes-hall]}}.

\bibitem{Geracie:2014nka}
M.~Geracie, D.~T. Son, C.~Wu, and S.-F. Wu, ``{Spacetime Symmetries of the
  Quantum Hall Effect},''
  \href{http://dx.doi.org/10.1103/PhysRevD.91.045030}{{\em Phys.Rev.}
  {\bfseries D91} (2015) 045030},
\href{http://arxiv.org/abs/1407.1252}{{\ttfamily arXiv:1407.1252
  [cond-mat.mes-hall]}}.

\bibitem{Jensen:2014aia}
K.~Jensen, ``{On the coupling of Galilean-invariant field theories to curved
  spacetime},''
\href{http://arxiv.org/abs/1408.6855}{{\ttfamily arXiv:1408.6855 [hep-th]}}.

\bibitem{Hartong:2014pma}
J.~Hartong, E.~Kiritsis, and N.~A. Obers, ``{Schr{\"o}dinger Invariance from
  Lifshitz Isometries in Holography and Field Theory},''
  \href{http://dx.doi.org/10.1103/PhysRevD.92.066003}{{\em Phys. Rev.}
  {\bfseries D92} (2015) 066003},
\href{http://arxiv.org/abs/1409.1522}{{\ttfamily arXiv:1409.1522 [hep-th]}}.

\bibitem{Hartong:2015wxa}
J.~Hartong, E.~Kiritsis, and N.~A. Obers, ``{Field Theory on Newton-Cartan
  Backgrounds and Symmetries of the Lifshitz Vacuum},''
  \href{http://dx.doi.org/10.1007/JHEP08(2015)006}{{\em JHEP} {\bfseries 08}
  (2015) 006},
\href{http://arxiv.org/abs/1502.00228}{{\ttfamily arXiv:1502.00228 [hep-th]}}.

\bibitem{Geracie:2015xfa}
M.~Geracie, K.~Prabhu, and M.~M. Roberts, ``{Fields and fluids on curved
  non-relativistic spacetimes},''
  \href{http://dx.doi.org/10.1007/JHEP08(2015)042}{{\em JHEP} {\bfseries 08}
  (2015) 042},
\href{http://arxiv.org/abs/1503.02680}{{\ttfamily arXiv:1503.02680 [hep-th]}}.

\bibitem{Gromov:2015fda}
A.~Gromov, K.~Jensen, and A.~G. Abanov, ``{Boundary effective action for
  quantum Hall states},''
  \href{http://dx.doi.org/10.1103/PhysRevLett.116.126802}{{\em Phys. Rev.
  Lett.} {\bfseries 116} no.~12, (2016) 126802},
\href{http://arxiv.org/abs/1506.07171}{{\ttfamily arXiv:1506.07171
  [cond-mat.str-el]}}.

\bibitem{Fuini:2015yva}
J.~F. Fuini, A.~Karch, and C.~F. Uhlemann, ``{Spinor fields in general
  Newton-Cartan backgrounds},''
  \href{http://dx.doi.org/10.1103/PhysRevD.92.125036}{{\em Phys. Rev.}
  {\bfseries D92} no.~12, (2015) 125036},
\href{http://arxiv.org/abs/1510.03852}{{\ttfamily arXiv:1510.03852 [hep-th]}}.

\bibitem{Bergshoeff:2015sic}
E.~Bergshoeff, J.~Rosseel, and T.~Zojer, ``{Non-relativistic fields from
  arbitrary contracting backgrounds},''
  \href{http://dx.doi.org/10.1088/0264-9381/33/17/175010}{{\em Class. Quant.
  Grav.} {\bfseries 33} no.~17, (2016) 175010},
\href{http://arxiv.org/abs/1512.06064}{{\ttfamily arXiv:1512.06064 [hep-th]}}.

\bibitem{Geracie:2016inm}
M.~Geracie, K.~Prabhu, and M.~M. Roberts, ``{Covariant effective action for a
  Galilean invariant quantum Hall system},''
  \href{http://dx.doi.org/10.1007/JHEP09(2016)092}{{\em JHEP} {\bfseries 09}
  (2016) 092},
\href{http://arxiv.org/abs/1603.08934}{{\ttfamily arXiv:1603.08934
  [cond-mat.mes-hall]}}.

\bibitem{Gromov:2017qeb}
A.~Gromov and D.~T. Son, ``{Bimetric Theory of Fractional Quantum Hall
  States},'' \href{http://dx.doi.org/10.1103/PhysRevX.7.041032}{{\em Phys.
  Rev.} {\bfseries X7} no.~4, (2017) 041032},
\href{http://arxiv.org/abs/1705.06739}{{\ttfamily arXiv:1705.06739
  [cond-mat.str-el]}}.

\bibitem{Arav:2016xjc}
I.~Arav, S.~Chapman, and Y.~Oz, ``{Non-Relativistic Scale Anomalies},''
  \href{http://dx.doi.org/10.1007/JHEP06(2016)158}{{\em JHEP} {\bfseries 06}
  (2016) 158},
\href{http://arxiv.org/abs/1601.06795}{{\ttfamily arXiv:1601.06795 [hep-th]}}.

\bibitem{Auzzi:2017wwc}
R.~Auzzi, S.~Baiguera, and G.~Nardelli, ``{Nonrelativistic trace and
  diffeomorphism anomalies in particle number background},''
  \href{http://dx.doi.org/10.1103/PhysRevD.97.085010}{{\em Phys. Rev.}
  {\bfseries D97} no.~8, (2018) 085010},
\href{http://arxiv.org/abs/1711.00910}{{\ttfamily arXiv:1711.00910 [hep-th]}}.

\bibitem{Bergshoeff:2015uaa}
E.~Bergshoeff, J.~Rosseel, and T.~Zojer, ``{Newton-Cartan (super)gravity as a
  non-relativistic limit},''
  \href{http://dx.doi.org/10.1088/0264-9381/32/20/205003}{{\em Class. Quant.
  Grav.} {\bfseries 32} no.~20, (2015) 205003},
\href{http://arxiv.org/abs/1505.02095}{{\ttfamily arXiv:1505.02095 [hep-th]}}.

\bibitem{Hartong:2015zia}
J.~Hartong and N.~A. Obers, ``{Ho{\v r}ava-Lifshitz gravity from dynamical
  Newton-Cartan geometry},''
  \href{http://dx.doi.org/10.1007/JHEP07(2015)155}{{\em JHEP} {\bfseries 07}
  (2015) 155},
\href{http://arxiv.org/abs/1504.07461}{{\ttfamily arXiv:1504.07461 [hep-th]}}.

\bibitem{Hartong:2015xda}
J.~Hartong, ``{Gauging the Carroll Algebra and Ultra-Relativistic Gravity},''
  \href{http://dx.doi.org/10.1007/JHEP08(2015)069}{{\em JHEP} {\bfseries 08}
  (2015) 069},
\href{http://arxiv.org/abs/1505.05011}{{\ttfamily arXiv:1505.05011 [hep-th]}}.

\bibitem{Afshar:2015aku}
H.~R. Afshar, E.~A. Bergshoeff, A.~Mehra, P.~Parekh, and B.~Rollier, ``{A
  Schr\"odinger approach to Newton-Cartan and Ho\v{r}ava-Lifshitz gravities},''
  \href{http://dx.doi.org/10.1007/JHEP04(2016)145}{{\em JHEP} {\bfseries 04}
  (2016) 145},
\href{http://arxiv.org/abs/1512.06277}{{\ttfamily arXiv:1512.06277 [hep-th]}}.

\bibitem{Bergshoeff:2016lwr}
E.~A. Bergshoeff and J.~Rosseel, ``{Three-Dimensional Extended Bargmann
  Supergravity},'' \href{http://dx.doi.org/10.1103/PhysRevLett.116.251601}{{\em
  Phys. Rev. Lett.} {\bfseries 116} no.~25, (2016) 251601},
\href{http://arxiv.org/abs/1604.08042}{{\ttfamily arXiv:1604.08042 [hep-th]}}.

\bibitem{Hartong:2016yrf}
J.~Hartong, Y.~Lei, and N.~A. Obers, ``{Nonrelativistic Chern-Simons theories
  and three-dimensional Ho{\v r}ava-Lifshitz gravity},''
  \href{http://dx.doi.org/10.1103/PhysRevD.94.065027}{{\em Phys. Rev.}
  {\bfseries D94} no.~6, (2016) 065027},
\href{http://arxiv.org/abs/1604.08054}{{\ttfamily arXiv:1604.08054 [hep-th]}}.

\bibitem{Bergshoeff:2017btm}
E.~Bergshoeff, J.~Gomis, B.~Rollier, J.~Rosseel, and T.~ter Veldhuis,
  ``{Carroll versus Galilei Gravity},''
  \href{http://dx.doi.org/10.1007/JHEP03(2017)165}{{\em JHEP} {\bfseries 03}
  (2017) 165},
\href{http://arxiv.org/abs/1701.06156}{{\ttfamily arXiv:1701.06156 [hep-th]}}.

\bibitem{VandenBleeken:2017rij}
D.~Van~den Bleeken, ``{Torsional Newton-Cartan gravity from the large c
  expansion of general relativity},''
  \href{http://dx.doi.org/10.1088/1361-6382/aa83d4}{{\em Class. Quant. Grav.}
  {\bfseries 34} no.~18, (2017) 185004},
\href{http://arxiv.org/abs/1703.03459}{{\ttfamily arXiv:1703.03459 [gr-qc]}}.

\bibitem{Bergshoeff:2017dqq}
E.~Bergshoeff, A.~Chatzistavrakidis, L.~Romano, and J.~Rosseel,
  ``{Newton-Cartan Gravity and Torsion},''
  \href{http://dx.doi.org/10.1007/JHEP10(2017)194}{{\em JHEP} {\bfseries 10}
  (2017) 194},
\href{http://arxiv.org/abs/1708.05414}{{\ttfamily arXiv:1708.05414 [hep-th]}}.

\bibitem{Andringa:2012uz}
R.~Andringa, E.~Bergshoeff, J.~Gomis, and M.~de~Roo, ``{'Stringy' Newton-Cartan
  Gravity},'' \href{http://dx.doi.org/10.1088/0264-9381/29/23/235020}{{\em
  Class.Quant.Grav.} {\bfseries 29} (2012) 235020},
\href{http://arxiv.org/abs/1206.5176}{{\ttfamily arXiv:1206.5176 [hep-th]}}.

\bibitem{Harmark:2017rpg}
T.~Harmark, J.~Hartong, and N.~A. Obers, ``{Nonrelativistic strings and limits
  of the AdS/CFT correspondence},''
  \href{http://dx.doi.org/10.1103/PhysRevD.96.086019}{{\em Phys. Rev.}
  {\bfseries D96} no.~8, (2017) 086019},
\href{http://arxiv.org/abs/1705.03535}{{\ttfamily arXiv:1705.03535 [hep-th]}}.

\bibitem{Andringa:2010it}
R.~Andringa, E.~Bergshoeff, S.~Panda, and M.~de~Roo, ``{Newtonian Gravity and
  the Bargmann Algebra},''
  \href{http://dx.doi.org/10.1088/0264-9381/28/10/105011}{{\em
  Class.Quant.Grav.} {\bfseries 28} (2011) 105011},
\href{http://arxiv.org/abs/1011.1145}{{\ttfamily arXiv:1011.1145 [hep-th]}}.

\bibitem{Bergshoeff:2014uea}
E.~A. Bergshoeff, J.~Hartong, and J.~Rosseel, ``{Torsional Newton-Cartan
  geometry and the Schr\"odinger algebra},''
  \href{http://dx.doi.org/10.1088/0264-9381/32/13/135017}{{\em Class. Quant.
  Grav.} {\bfseries 32} no.~13, (2015) 135017},
\href{http://arxiv.org/abs/1409.5555}{{\ttfamily arXiv:1409.5555 [hep-th]}}.

\bibitem{Festuccia:2016awg}
G.~Festuccia, D.~Hansen, J.~Hartong, and N.~A. Obers, ``{Torsional
  Newton-Cartan Geometry from the Noether Procedure},''
  \href{http://dx.doi.org/10.1103/PhysRevD.94.105023}{{\em Phys. Rev.}
  {\bfseries D94} no.~10, (2016) 105023},
\href{http://arxiv.org/abs/1607.01926}{{\ttfamily arXiv:1607.01926 [hep-th]}}.

\bibitem{Griffin:2012qx}
T.~Griffin, P.~Ho\v{r}ava, and C.~M. Melby-Thompson, ``{Lifshitz Gravity for
  Lifshitz Holography},''
  \href{http://dx.doi.org/10.1103/PhysRevLett.110.081602}{{\em Phys.Rev.Lett.}
  {\bfseries 110} no.~8, (2013) 081602},
\href{http://arxiv.org/abs/1211.4872}{{\ttfamily arXiv:1211.4872 [hep-th]}}.

\bibitem{Janiszewski:2012nf}
S.~Janiszewski and A.~Karch, ``{String Theory Embeddings of Nonrelativistic
  Field Theories and Their Holographic Ho\v{r}ava Gravity Duals},''
  \href{http://dx.doi.org/10.1103/PhysRevLett.110.081601}{{\em Phys.Rev.Lett.}
  {\bfseries 110} no.~8, (2013) 081601},
\href{http://arxiv.org/abs/1211.0010}{{\ttfamily arXiv:1211.0010 [hep-th]}}.

\bibitem{Ryu:2006bv}
S.~Ryu and T.~Takayanagi, ``{Holographic derivation of entanglement entropy
  from AdS/CFT},'' \href{http://dx.doi.org/10.1103/PhysRevLett.96.181602}{{\em
  Phys. Rev. Lett.} {\bfseries 96} (2006) 181602},
\href{http://arxiv.org/abs/hep-th/0603001}{{\ttfamily arXiv:hep-th/0603001
  [hep-th]}}.

\bibitem{Ammon:2013hba}
M.~Ammon, A.~Castro, and N.~Iqbal, ``{Wilson Lines and Entanglement Entropy in
  Higher Spin Gravity},'' \href{http://dx.doi.org/10.1007/JHEP10(2013)110}{{\em
  JHEP} {\bfseries 10} (2013) 110},
\href{http://arxiv.org/abs/1306.4338}{{\ttfamily arXiv:1306.4338 [hep-th]}}.

\bibitem{deBoer:2013vca}
J.~de~Boer and J.~I. Jottar, ``{Entanglement Entropy and Higher Spin Holography
  in AdS$_3$},'' \href{http://dx.doi.org/10.1007/JHEP04(2014)089}{{\em JHEP}
  {\bfseries 04} (2014) 089},
\href{http://arxiv.org/abs/1306.4347}{{\ttfamily arXiv:1306.4347 [hep-th]}}.

\bibitem{Castro:2015csg}
A.~Castro, D.~M. Hofman, and N.~Iqbal, ``{Entanglement Entropy in Warped
  Conformal Field Theories},''
  \href{http://dx.doi.org/10.1007/JHEP02(2016)033}{{\em JHEP} {\bfseries 02}
  (2016) 033},
\href{http://arxiv.org/abs/1511.00707}{{\ttfamily arXiv:1511.00707 [hep-th]}}.

\bibitem{Song:2016pwx}
W.~Song, Q.~Wen, and J.~Xu, ``{Generalized Gravitational Entropy for Warped
  Anti--de Sitter Space},''
  \href{http://dx.doi.org/10.1103/PhysRevLett.117.011602}{{\em Phys. Rev.
  Lett.} {\bfseries 117} no.~1, (2016) 011602},
\href{http://arxiv.org/abs/1601.02634}{{\ttfamily arXiv:1601.02634 [hep-th]}}.

\bibitem{Song:2016gtd}
W.~Song, Q.~Wen, and J.~Xu, ``{Modifications to Holographic Entanglement
  Entropy in Warped CFT},''
  \href{http://dx.doi.org/10.1007/JHEP02(2017)067}{{\em JHEP} {\bfseries 02}
  (2017) 067},
\href{http://arxiv.org/abs/1610.00727}{{\ttfamily arXiv:1610.00727 [hep-th]}}.

\bibitem{Jiang:2017ecm}
H.~Jiang, W.~Song, and Q.~Wen, ``{Entanglement Entropy in Flat Holography},''
  \href{http://dx.doi.org/10.1007/JHEP07(2017)142}{{\em JHEP} {\bfseries 07}
  (2017) 142},
\href{http://arxiv.org/abs/1706.07552}{{\ttfamily arXiv:1706.07552 [hep-th]}}.

\bibitem{Taylor:2015glc}
M.~Taylor, ``{Lifshitz holography},'' {\em Class. Quant. Grav.} {\bfseries 33}
  no.~3, (2016) 033001,
\href{http://arxiv.org/abs/1512.03554}{{\ttfamily arXiv:1512.03554 [hep-th]}}.

\bibitem{SchaferNameki:2009xr}
S.~Schafer-Nameki, M.~Yamazaki, and K.~Yoshida, ``{Coset Construction for Duals
  of Non-relativistic CFTs},''
  \href{http://dx.doi.org/10.1088/1126-6708/2009/05/038}{{\em JHEP} {\bfseries
  0905} (2009) 038},
\href{http://arxiv.org/abs/0903.4245}{{\ttfamily arXiv:0903.4245 [hep-th]}}.

\bibitem{Jottar:2010vp}
J.~I. Jottar, R.~G. Leigh, D.~Minic, and L.~A. Pando~Zayas, ``{Aging and
  Holography},'' \href{http://dx.doi.org/10.1007/JHEP11(2010)034}{{\em JHEP}
  {\bfseries 11} (2010) 034},
\href{http://arxiv.org/abs/1004.3752}{{\ttfamily arXiv:1004.3752 [hep-th]}}.

\bibitem{Bagchi:2010xw}
A.~Bagchi and A.~Kundu, ``{Metrics with Galilean Conformal Isometry},''
  \href{http://dx.doi.org/10.1103/PhysRevD.83.066018}{{\em Phys. Rev.}
  {\bfseries D83} (2011) 066018},
\href{http://arxiv.org/abs/1011.4999}{{\ttfamily arXiv:1011.4999 [hep-th]}}.

\bibitem{Duval:2012qr}
C.~Duval and S.~Lazzarini, ``{Schr\"odinger Manifolds},''
  \href{http://dx.doi.org/10.1088/1751-8113/45/39/395203}{{\em J. Phys.}
  {\bfseries A45} (2012) 395203},
\href{http://arxiv.org/abs/1201.0683}{{\ttfamily arXiv:1201.0683 [math-ph]}}.

\bibitem{Nappi:1993ie}
C.~R. Nappi and E.~Witten, ``{A WZW model based on a nonsemisimple group},''
  \href{http://dx.doi.org/10.1103/PhysRevLett.71.3751}{{\em Phys. Rev. Lett.}
  {\bfseries 71} (1993) 3751--3753},
\href{http://arxiv.org/abs/hep-th/9310112}{{\ttfamily arXiv:hep-th/9310112
  [hep-th]}}.

\bibitem{Kachru:2008yh}
S.~Kachru, X.~Liu, and M.~Mulligan, ``{Gravity duals of Lifshitz-like fixed
  points},'' \href{http://dx.doi.org/10.1103/PhysRevD.78.106005}{{\em Phys.
  Rev.} {\bfseries D78} (2008) 106005},
\href{http://arxiv.org/abs/0808.1725}{{\ttfamily arXiv:0808.1725 [hep-th]}}.

\bibitem{Taylor:2008tg}
M.~Taylor, ``{Non-relativistic holography},''
\href{http://arxiv.org/abs/0812.0530}{{\ttfamily arXiv:0812.0530 [hep-th]}}.

\bibitem{Son:2008ye}
D.~T. Son, ``{Toward an AdS/cold atoms correspondence: A Geometric realization
  of the Schrodinger symmetry},''
  \href{http://dx.doi.org/10.1103/PhysRevD.78.046003}{{\em Phys. Rev.}
  {\bfseries D78} (2008) 046003},
\href{http://arxiv.org/abs/0804.3972}{{\ttfamily arXiv:0804.3972 [hep-th]}}.

\bibitem{Adams:2008wt}
A.~Adams, K.~Balasubramanian, and J.~McGreevy, ``{Hot Spacetimes for Cold
  Atoms},'' \href{http://dx.doi.org/10.1088/1126-6708/2008/11/059}{{\em JHEP}
  {\bfseries 11} (2008) 059},
\href{http://arxiv.org/abs/0807.1111}{{\ttfamily arXiv:0807.1111 [hep-th]}}.

\bibitem{Keeler:2014lia}
C.~Keeler, G.~Knodel, and J.~T. Liu, ``{Hidden horizons in non-relativistic
  AdS/CFT},'' \href{http://dx.doi.org/10.1007/JHEP08(2014)024}{{\em JHEP}
  {\bfseries 08} (2014) 024},
\href{http://arxiv.org/abs/1404.4877}{{\ttfamily arXiv:1404.4877 [hep-th]}}.

\bibitem{Keeler:2013msa}
C.~Keeler, G.~Knodel, and J.~T. Liu, ``{What do non-relativistic CFTs tell us
  about Lifshitz spacetimes?},''
  \href{http://dx.doi.org/10.1007/JHEP01(2014)062}{{\em JHEP} {\bfseries 1401}
  (2014) 062},
\href{http://arxiv.org/abs/1308.5689}{{\ttfamily arXiv:1308.5689 [hep-th]}}.

\bibitem{Keeler:2015afa}
C.~Keeler, G.~Knodel, J.~T. Liu, and K.~Sun, ``{Universal features of Lifshitz
  Green's functions from holography},''
  \href{http://dx.doi.org/10.1007/JHEP08(2015)057}{{\em JHEP} {\bfseries 08}
  (2015) 057},
\href{http://arxiv.org/abs/1505.07830}{{\ttfamily arXiv:1505.07830 [hep-th]}}.

\bibitem{Gentle:2015cfp}
S.~A. Gentle and C.~Keeler, ``{On the reconstruction of Lifshitz spacetimes},''
  \href{http://dx.doi.org/10.1007/JHEP03(2016)195}{{\em JHEP} {\bfseries 03}
  (2016) 195},
\href{http://arxiv.org/abs/1512.04538}{{\ttfamily arXiv:1512.04538 [hep-th]}}.

\bibitem{Brauner:2014jaa}
T.~Brauner, S.~Endlich, A.~Monin, and R.~Penco, ``{General coordinate
  invariance in quantum many-body systems},''
  \href{http://dx.doi.org/10.1103/PhysRevD.90.105016}{{\em Phys.Rev.}
  {\bfseries D90} no.~10, (2014) 105016},
\href{http://arxiv.org/abs/1407.7730}{{\ttfamily arXiv:1407.7730 [hep-th]}}.

\bibitem{Geracie:2015dea}
M.~Geracie, K.~Prabhu, and M.~M. Roberts, ``{Curved non-relativistic
  spacetimes, Newtonian gravitation and massive matter},''
  \href{http://dx.doi.org/10.1063/1.4932967}{{\em J. Math. Phys.} {\bfseries
  56} no.~10, (2015) 103505},
\href{http://arxiv.org/abs/1503.02682}{{\ttfamily arXiv:1503.02682 [hep-th]}}.

\bibitem{Bekaert:2015xua}
X.~Bekaert and K.~Morand, ``{Connections and dynamical trajectories in
  generalised Newton-Cartan gravity II. An ambient perspective},''
  \href{http://dx.doi.org/10.1063/1.5030328}{{\em J. Math. Phys.} {\bfseries
  59} (2018) 072503},
\href{http://arxiv.org/abs/1505.03739}{{\ttfamily arXiv:1505.03739 [hep-th]}}.

\bibitem{Karananas:2016hrm}
G.~K. Karananas and A.~Monin, ``{Gauging nonrelativistic field theories using
  the coset construction},''
  \href{http://dx.doi.org/10.1103/PhysRevD.93.064069}{{\em Phys. Rev.}
  {\bfseries D93} (2016) 064069},
\href{http://arxiv.org/abs/1601.03046}{{\ttfamily arXiv:1601.03046 [hep-th]}}.

\bibitem{Duval:2011mi}
C.~Duval and P.~Horvathy, ``{Conformal Galilei groups, Veronese curves, and
  Newton-Hooke spacetimes},''
  \href{http://dx.doi.org/10.1088/1751-8113/44/33/335203}{{\em J. Phys.}
  {\bfseries A44} (2011) 335203},
\href{http://arxiv.org/abs/1104.1502}{{\ttfamily arXiv:1104.1502 [hep-th]}}.

\bibitem{Duval:2016tzi}
C.~Duval, G.~Gibbons, and P.~Horvathy, ``{Conformal and projective symmetries
  in Newtonian cosmology},''
  \href{http://dx.doi.org/10.1016/j.geomphys.2016.11.012}{{\em J. Geom. Phys.}
  {\bfseries 112} (2017) 197--209},
\href{http://arxiv.org/abs/1605.00231}{{\ttfamily arXiv:1605.00231 [gr-qc]}}.

\bibitem{Bekaert:2014bwa}
X.~Bekaert and K.~Morand, ``{Connections and dynamical trajectories in
  generalised Newton-Cartan gravity I. An intrinsic view},''
  \href{http://dx.doi.org/10.1063/1.4937445}{{\em J. Math. Phys.} {\bfseries
  57} no.~2, (2016) 022507},
\href{http://arxiv.org/abs/1412.8212}{{\ttfamily arXiv:1412.8212 [hep-th]}}.

\bibitem{Duval:1993pe}
C.~Duval, ``{On Galileian isometries},''
  \href{http://dx.doi.org/10.1088/0264-9381/10/11/006}{{\em Class.Quant.Grav.}
  {\bfseries 10} (1993) 2217--2222},
\href{http://arxiv.org/abs/0903.1641}{{\ttfamily arXiv:0903.1641 [math-ph]}}.

\bibitem{Trumper:1983}
M.~Tr{\"u}mper, ``Lagrangian mechanics and the geometry of configuration
  spacetime,'' \href{http://dx.doi.org/10.1016/0003-4916(83)90305-6}{{\em
  Annals Phys.} {\bfseries 149} no.~1, (1983) 203--233}.

\bibitem{Duval:1984cj}
C.~Duval, G.~Burdet, H.~Kunzle, and M.~Perrin, ``{Bargmann Structures and
  Newton-Cartan Theory},''
\href{http://dx.doi.org/10.1103/PhysRevD.31.1841}{{\em Phys.Rev.} {\bfseries
  D31} (1985) 1841}.

\bibitem{Duval:1990hj}
C.~Duval, G.~W. Gibbons, and P.~Horvathy, ``{Celestial mechanics, conformal
  structures and gravitational waves},''
  \href{http://dx.doi.org/10.1103/PhysRevD.43.3907}{{\em Phys.Rev.} {\bfseries
  D43} (1991) 3907--3922},
\href{http://arxiv.org/abs/hep-th/0512188}{{\ttfamily arXiv:hep-th/0512188
  [hep-th]}}.

\bibitem{Julia:1994bs}
B.~Julia and H.~Nicolai, ``{Null Killing vector dimensional reduction and
  Galilean geometrodynamics},''
  \href{http://dx.doi.org/10.1016/0550-3213(94)00584-2}{{\em Nucl.Phys.}
  {\bfseries B439} (1995) 291--326},
\href{http://arxiv.org/abs/hep-th/9412002}{{\ttfamily arXiv:hep-th/9412002
  [hep-th]}}.

\bibitem{Bacry:1968zf}
H.~Bacry and J.~Levy-Leblond, ``{Possible kinematics},''
\href{http://dx.doi.org/10.1063/1.1664490}{{\em J.Math.Phys.} {\bfseries 9}
  (1968) 1605--1614}.

\bibitem{Ballesteros:1994}
A.~Ballesteros, F.~J. Herranz, M.~A.~D. Olmo, and M.~Santander, ``Quantum (2+1)
  kinematical algebras: a global approach,''
  \href{http://dx.doi.org/10.1088/0305-4470/27/4/021/meta}{{\em Journal of
  Physics A: Mathematical and General} {\bfseries 27} no.~4, (1994) 1283}.

\bibitem{Herranz:2002}
F.~J. Herranz and M.~Santander, ``Conformal symmetries of spacetimes,''
  \href{http://dx.doi.org/10.1088/0305-4470/35/31/306}{{\em Journal of Physics
  A: Mathematical and General} {\bfseries 35} no.~31, (2002) 6601}.

\bibitem{Gibbons:2003rv}
G.~W. Gibbons and C.~E. Patricot, ``{Newton-Hooke space-times, Hpp waves and
  the cosmological constant},''
  \href{http://dx.doi.org/10.1088/0264-9381/20/23/016}{{\em Class. Quant.
  Grav.} {\bfseries 20} (2003) 5225},
\href{http://arxiv.org/abs/hep-th/0308200}{{\ttfamily arXiv:hep-th/0308200
  [hep-th]}}.

\bibitem{Bergshoeff:2014gja}
E.~Bergshoeff, J.~Gomis, M.~Kovacevic, L.~Parra, J.~Rosseel, {\em et~al.},
  ``{Nonrelativistic superparticle in a curved background},''
  \href{http://dx.doi.org/10.1103/PhysRevD.90.065006}{{\em Phys.Rev.}
  {\bfseries D90} no.~6, (2014) 065006},
\href{http://arxiv.org/abs/1406.7286}{{\ttfamily arXiv:1406.7286 [hep-th]}}.

\bibitem{Figueroa-OFarrill:2017sfs}
J.~Figueroa-O'Farrill, ``{Classification of kinematical Lie algebras},''
\href{http://arxiv.org/abs/1711.05676}{{\ttfamily arXiv:1711.05676 [hep-th]}}.

\bibitem{Figueroa-OFarrill:2017ycu}
J.~M. Figueroa-O'Farrill, ``{Kinematical Lie algebras via deformation
  theory},'' \href{http://dx.doi.org/10.1063/1.5016288}{{\em J. Math. Phys.}
  {\bfseries 59} no.~6, (2018) 061701},
\href{http://arxiv.org/abs/1711.06111}{{\ttfamily arXiv:1711.06111 [hep-th]}}.

\bibitem{Figueroa-OFarrill:2017tcy}
J.~M. Figueroa-O'Farrill, ``{Higher-dimensional kinematical Lie algebras via
  deformation theory},'' \href{http://dx.doi.org/10.1063/1.5016616}{{\em J.
  Math. Phys.} {\bfseries 59} no.~6, (2018) 061702},
\href{http://arxiv.org/abs/1711.07363}{{\ttfamily arXiv:1711.07363 [hep-th]}}.

\bibitem{Hartong:2017bwq}
J.~Hartong, Y.~Lei, N.~A. Obers, and G.~Oling, ``{Zooming in on
  AdS$_{3}$/CFT$_{2}$ near a BPS bound},''
  \href{http://dx.doi.org/10.1007/JHEP05(2018)016}{{\em JHEP} {\bfseries 05}
  (2018) 016},
\href{http://arxiv.org/abs/1712.05794}{{\ttfamily arXiv:1712.05794 [hep-th]}}.

\bibitem{Gomis:2005pg}
J.~Gomis, J.~Gomis, and K.~Kamimura, ``{Non-relativistic superstrings: A New
  soluble sector of AdS${}_5 \times S^5$},''
  \href{http://dx.doi.org/10.1088/1126-6708/2005/12/024}{{\em JHEP} {\bfseries
  12} (2005) 024},
\href{http://arxiv.org/abs/hep-th/0507036}{{\ttfamily arXiv:hep-th/0507036
  [hep-th]}}.

\bibitem{Brugues:2006yd}
J.~Brugues, J.~Gomis, and K.~Kamimura, ``{Newton-Hooke algebras,
  non-relativistic branes and generalized pp-wave metrics},''
  \href{http://dx.doi.org/10.1103/PhysRevD.73.085011}{{\em Phys. Rev.}
  {\bfseries D73} (2006) 085011},
\href{http://arxiv.org/abs/hep-th/0603023}{{\ttfamily arXiv:hep-th/0603023
  [hep-th]}}.

\bibitem{Hason:2017flq}
I.~Hason, ``{Triviality of Entanglement Entropy in the Galilean Vacuum},''
  \href{http://dx.doi.org/10.1016/j.physletb.2018.02.064}{{\em Phys. Lett.}
  {\bfseries B780} (2018) 149--151},
\href{http://arxiv.org/abs/1708.08303}{{\ttfamily arXiv:1708.08303 [hep-th]}}.

\bibitem{Hartong:2016nyx}
J.~Hartong, N.~A. Obers, and M.~Sanchioni, ``{Lifshitz Hydrodynamics from
  Lifshitz Black Branes with Linear Momentum},''
  \href{http://dx.doi.org/10.1007/JHEP10(2016)120}{{\em JHEP} {\bfseries 10}
  (2016) 120},
\href{http://arxiv.org/abs/1606.09543}{{\ttfamily arXiv:1606.09543 [hep-th]}}.

\bibitem{MohammadiMozaffar:2017nri}
M.~R. Mohammadi~Mozaffar and A.~Mollabashi, ``{Entanglement in Lifshitz-type
  Quantum Field Theories},''
  \href{http://dx.doi.org/10.1007/JHEP07(2017)120}{{\em JHEP} {\bfseries 07}
  (2017) 120},
\href{http://arxiv.org/abs/1705.00483}{{\ttfamily arXiv:1705.00483 [hep-th]}}.

\bibitem{He:2017wla}
T.~He, J.~M. Magan, and S.~Vandoren, ``{Entanglement Entropy in Lifshitz
  Theories},'' \href{http://dx.doi.org/10.21468/SciPostPhys.3.5.034}{{\em
  SciPost Phys.} {\bfseries 3} no.~5, (2017) 034},
\href{http://arxiv.org/abs/1705.01147}{{\ttfamily arXiv:1705.01147 [hep-th]}}.

\bibitem{Gentle:2017ywk}
S.~A. Gentle and S.~Vandoren, ``{Lifshitz entanglement entropy from holographic
  cMERA},'' \href{http://dx.doi.org/10.1007/JHEP07(2018)013}{{\em JHEP}
  {\bfseries 07} (2018) 013},
\href{http://arxiv.org/abs/1711.11509}{{\ttfamily arXiv:1711.11509 [hep-th]}}.

\end{thebibliography}
\end{document}